\documentclass[aps,prb,twocolumn,groupedaddress,showpacs,floatfix]{revtex4}

\usepackage{graphicx}

\begin{document}

\title{Thermoelectric effects in a strongly correlated  model for Na$_x$CoO$_2$}

\author{Michael R. Peterson}
\email{peterson@physics.ucsc.edu}
\affiliation{Physics Department, University of California,  Santa Cruz, CA  95064}
\author{B. Sriram Shastry}
\affiliation{Physics Department, University of California,  Santa Cruz, CA  95064}
\author{Jan O. Haerter}
\affiliation{Physics Department, University of California,  Santa Cruz, CA  95064}

\date{\today}

\begin{abstract}
Thermal response functions of strongly correlated electron systems are
of appreciable interest to the larger scientific community both theoretically 
and technologically.  Here we focus on the infinitely correlated
$t$-$J$ model on a geometrically frustrated two-dimensional triangular lattice.  
Using exact diagonalization on a finite sized system
we calculate the dynamical thermal response functions in order to determine 
the thermopower, Lorenz number, and dimensionless figure of merit.  The dynamical 
thermal response functions is compared to the infinite frequency limit and shown 
to be very weak functions of frequency, hence, establishing the 
validity of the high frequency formalism recently proposed 
by Shastry\cite{shastry_1,shastry_2,shastry_3} for the thermopower, Lorenz number, and 
the dimensionless figure of merit.  Further, the thermopower is 
demonstrated to have a low to mid temperature enhancement when the 
sign of the hopping parameter $t$ is switched from positive to negative 
for the geometrically frustrated lattice\cite{ramirez} considered.  
\end{abstract}


\maketitle

\section{Introduction}

There is current interest in the physics, as well as industrial and engineering
communities regarding thermoelectrics of strongly correlated electron 
systems.  This interest 
has been recently revived by the demonstration of the unexpectedly high 
thermopower seen in the very interesting material 
sodium cobalt oxide\cite{nco_terasaki,nco_general,ong_nature}(NCO).  
Theoretically, thermoelectrics 
have been a long standing problem in physics especially when concerned 
with strongly correlated systems which are not amenable to 
perturbative treatments.  

Essentially there are two standard theoretical approaches concerning the problem 
of thermoelectrics, cf. Ref.~\onlinecite{ziman,kubo,mahan,mahan_1}.  The first makes use of 
Boltzmann theory often complimented with standard Fermi liquid theory.  
This methodology is reliable 
for weakly coupled problems where long lived quasiparticles 
remain well defined and where perturbation theory remains valid.  The second 
approach is to use the full rigor of the Kubo formalism which is
valid for all situations but whose dynamical character make 
it unyielding and difficult to make any real progress, 
especially for strongly correlated systems.

Recently, Shastry\cite{shastry_1,shastry_2,shastry_3} has proposed a third method which 
handles the strong electron interactions with the respect which they deserve 
while avoiding the complexity of the full dynamics of the Kubo formalism.  Very briefly, 
this methodology considers the ultimate quantities one is usually interested 
in when calculating conductivities.  Often there is more interest in certain combinations of 
conductivities, which form more experimentally accessible quantities (such 
as the thermopower or Seebeck coefficient, Lorenz number, dimensionless 
figure of merit, the Hall coefficient, etc.), than in the conductivities (electrical, 
thermal, etc.) themselves.  The basic proposal is that certain combinations of conductivities 
have weak dynamical character (weak frequency dependence), and thus 
lend themselves to a high frequency expansion.  The upshot of this expansion is 
that it yields formulas that are much simpler, although non-trivial, than the Kubo 
formulas and yet the interactions are fully respected compared to the 
usual approximations which risk missing important effects.  

Nearly 15 years previously, this basic high 
frequency expansion methodology was originally employed  
by Shastry, Shraiman, and Singh\cite{sss} to calculate the Hall 
coefficient (at high temperatures)
for a strongly correlated electron model ($t$-$J$ model) with success.  
In the last year, the present authors\cite{cw-prl,cw-prb} have 
applied this high frequency expansion to calculate the 
Hall coefficient and thermopower for the very interesting NCO 
system explaining in both quantitative and qualitative detail the 
physics of the so-called Curie-Weiss metal for Na$_x$CoO$_2$ at 
electron doping $x\sim0.7$.  This Curie-Weiss metal displays 
behavior that is an interesting hybrid between those of insulating and 
metallic systems.  The high frequency formalism allowed the investigation 
of this complicated system successfully by incorporating the 
important effects of interactions.

Interestingly, Shastry, via the high frequency expansion, 
was able to predict a low to mid temperature thermopower enhancement due to a change 
in sign of the hopping parameter $t$ of the $t$-$J$ model for 
a geometrically frustrated\cite{ramirez} two-dimensional triangular lattice\cite{shastry_1,shastry_2}.  
This corresponds to a fiduciary hole doped CoO$_2$ layer of NCO that has yet to 
be realized experimentally.  The lattice topology plays a crucial role in this enhancement as it 
owes its existence primarily to electron transport and is not thermodynamic 
or entropic in origin.  This prediction was put on firmer footing by the present 
authors in Ref.~\onlinecite{cw-prl} concerning the Curie-Weiss 
metallic phase of NCO which itself 
has an underlying two-dimensional triangular lattice.  We emphasize that we 
work with a hole doped system ($0\leq n\leq1$, $n$ electron density) and 
in order to compare with experiments on NCO, we perform a suitable 
particle-hole transformation\cite{shastry_1,shastry_2,cw-prl,cw-prb}.

In this work, we establish the validity and accuracy of the 
high frequency formalism for the thermopower, Lorenz number, and figure 
of merit for the strongly correlated electron $t$-$J$ model on a 
two-dimensional triangular lattice.  This is accomplished 
by comparing the high frequency expressions with those obtained 
via the full Kubo formalism.  This comparison for a strongly 
correlated system, such as our model, is only possible through 
numerical exact diagonalization of a relatively small system 
($\mathcal{L}=12$ site lattice).  However, we feel that our results 
provide a much desired and important benchmark for the 
high frequency formalism establishing its effectiveness and usefulness.

Furthermore, the $t$-$J$ model is generally  
representative of strongly correlated electron models so 
our results should be applicable to other strong correlation 
models such as the Hubbard model for large $U$.  While the 
geometrically frustrated triangular lattice provides an interesting 
enhancement of the thermopower for intermediate temperatures, 
the general validity of our results should obtain for other lattice 
topologies (frustrated and non-frustrated).

The plan of this paper is as follows:  In section~\ref{sec-model} we describe 
the details of our model and the exact diagonalization used, section~\ref{sec-kubo} 
generally quotes the Kubo formulas for the considered conductivities and 
the high frequency formulas from Shastry\cite{shastry_1,shastry_2}.  In 
sections~\ref{sec-s}-\ref{sec-zt} we report results for
the thermopower (for both positive and negative hopping $t$), the Lorenz 
number, and the figure of merit (both for positive hopping $t>0$), respectively.  
Section~\ref{sec-conc} concludes 
while some formulas are given for completeness in the appendix.

\section{$t$-$J$ model and Diagonalization}\label{sec-model}

As mentioned above we study the $t$-$J$ model Hamiltonian 
which describes a strongly correlated hole doped Mott insulator.  The Hamiltonian is
\begin{eqnarray}
\hat{H} =  -\sum_{\vec r\vec\eta\sigma}t(\vec\eta)\tilde
c^{\dagger}_{\vec r+\vec\eta\sigma} \tilde c_{\vec r\sigma}   +
\frac{1}{2}\sum_{\vec r\vec \eta}J(\vec \eta)\vec{S}_{\vec r}\cdot
\vec{S}_{\vec r + \vec \eta}\;,
\end{eqnarray}
where $\tilde c^{\dagger}_{\vec
r\sigma}(\tilde c_{\vec r\sigma})=\hat{P}_Gc^{\dagger}_{\vec
r\sigma}(c_{\vec r\sigma})\hat{P}_G$ are Gutzwiller projected 
fermion creation(destruction) operators where the projection operator
$\hat{P}_G$ projects out all doubly occupied lattice sites.  The lattice vector 
$\vec\eta$ connects nearest neighbors which are coupled (with strength $J(\vec\eta)$) via their 
spin degree of freedom ($\vec{S}_{\vec r}$ is the three-component 
spin operator).  For simplicity we take the hopping $t(\vec\eta)=t$ and 
spin $J(\vec\eta)=J$ coupling parameters to be constants and the 
lattice constant has been set to unity.

In this work we also aim to apply our calculations to the 
experimental system of NCO which is electron-doped 
and has been previously modeled using the $t$-$J$ model\cite{kumar,baskaran,lee,cw-prl}.
We use the symmetry of the Hubbard model 
with regard to half filling to map our system 
to NCO, i.e., we apply the replacement rules $t\rightarrow -t$, 
doping $x=|1-n|$, and $q_e\rightarrow-q_e$ where $n$ is electron 
density per site and $q_e=-|e|$ is the electron charge.  At this 
point we will abandon referencing particular systems by the 
electron density and instead reference them by the doping $x\equiv n-1$.

\begin{figure}[t]
\begin{center}
\includegraphics[width=7.5cm]{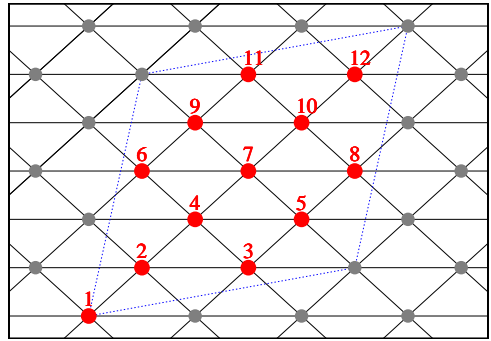}
\end{center}
\caption{$\mathcal{L}=12$ site torus geometry used throughout this work.  We 
diagonalize the Hamiltonian at all densities of this lattice for the $t$-$J$ model, however, as 
explained in the text we label the electron densities in terms of hole doping $x$.}
\label{L12}
\end{figure}

Thermodynamics will be considered within the canonical ensemble and 
considering that the goal of this paper is to calculate full thermodynamic 
Kubo formulas it is a prerequisite that we obtain the full 
eigenspectrum of the system in order to calculate all relevant current 
matrix elements.  Therefore we make progress through the exact numerical diagonalization 
of a finite system.  At this level of study various approximations 
such as the finite temperature Lanczos method\cite{ftl}, dynamical 
mean field theory\cite{dmft}, etc., could perhaps muddy the issue 
of comparing the high frequency expansion of various thermoelectric 
properties to their full Kubo formulations and, hence, will not 
be considered.  Further, we are interested in the behavior of 
the system over a whole range of densities and do not wish to 
confuse the results by considering a smattering of small system sizes along with 
their particular inherent and unavoidable finite size effects.  Therefore, 
we find that the largest two-dimensional lattice that we can fully diagonalize in 
a satisfactory way is a $\mathcal{L}=12$ site toroidal lattice, c.f. 
Fig~\ref{L12}.  Another reason for using this lattice is that it 
was used extensively in the present authors' previous works\cite{cw-prl,cw-prb} on the 
two-dimensional triangular lattice $t$-$J$ model applied to NCO.

To reduce the computational demand of the exact diagonalization to a more 
manageable size we employ a number of symmetries.  Firstly, we consider 
only the largest $S_z$ sector of the full Hilbert space, i.e., the smallest 
$|S_z|$ subspace which is $|S_z|=0(\hbar/2)$ for even(odd) number 
of electrons.  This sector of the full Hilbert space dominates the 
physics so this ``approximation'' is as good as exact\cite{to-be}.  The most 
useful symmetry employed is translational invariance which 
essentially reduces the Hilbert space dimension by a factor of 
$\mathcal{L}$.  Our largest Hilbert space dimension occurs for 
$x=1/3$ corresponding to $34650$ states in the $|S_z|=0$ subspace and after 
applying translational symmetry we need only diagonalize matrices 
of dimension $\sim 2900$.  While this matrix dimension is not particularly 
huge in the realm of matrix diagonalization we must consider a 
double sum over these $\sim 2900$ states to calculate the Kubo formulas.  This double 
sum is quite time consuming and limits our abilities to consider larger 
lattices and, in fact, it limits our abilities to consider all 
doping values $x$ on the chosen $\mathcal{L}=12$ site lattice used here.  
The dopings where we can calculate the full Kubo formula is 
limited to $x>0.5$ and $x<0.2$.

\section{Dynamical Thermal Response Functions}\label{sec-kubo}

In this section we quote the formulas for the Kubo linear response for 
thermoelectrics following very closely the work of 
Shastry\cite{shastry_1,shastry_2,shastry_3}.  In particular we are interested in 
the electrical $\sigma(\omega,T)$,  thermoelectrical
$\gamma(\omega,T)$, and the thermal $\kappa(\omega,T)$ conductivities, respectively.  
In terms of these conductivities the ultimate goal of this work are 
commonly measured physical quantities of interest such as the thermopower $S$, 
the Lorenz number $L$, and the dimensionless figure
of merit $ZT$ commonly defined\cite{ziman,mahan,mahan_1} as
\begin{eqnarray}
S(\omega,T)=\frac{\gamma(\omega,T)}{\sigma(\omega,T)}\;,
\label{thermo}
\end{eqnarray}
\begin{eqnarray}
L(\omega,T)=\frac{\kappa(\omega,T)}{T\sigma(\omega,T)}-\{S(\omega,T)\}^2\;,
\label{lorenz}
\end{eqnarray}
and
\begin{eqnarray}\label{fom}
Z(\omega,T)T=\frac{\{S(\omega,T)\}^2}{L(\omega,T)}\;.
\end{eqnarray}

These conductivities all have familiar Kubo formulas which, in the Lehmann representation, are written as
\begin{widetext}
\begin{eqnarray}\label{kubo-sig}
\sigma(\omega_c,T)=\frac{i}{\hbar\omega_c \mathcal{L}}\left[ \langle
\hat\tau_{xx}\rangle+\frac{\hbar}{\mathcal{Z}}\sum_{n,m}\frac{e^{-\beta\varepsilon_n}-
e^{-\beta\varepsilon_m}}{\varepsilon_n-\varepsilon_m+\hbar\omega_c}|\langle
n|\hat{J}_x|m\rangle|^2\right]\;,
\end{eqnarray}
\begin{eqnarray}\label{kubo-gam}
\gamma(\omega_c,T)=\frac{i}{\hbar\omega_c T \mathcal{L}}\left[ \langle
\hat\Phi_{xx}\rangle+\frac{\hbar}{\mathcal{Z}}
\sum_{n,m}\frac{e^{-\beta\varepsilon_n}-e^{-\beta\varepsilon_m}}{\varepsilon_n-\varepsilon_m+
\hbar\omega_c}\langle
n|\hat{J}_x|m\rangle \langle m|\hat{J}^Q_x|n\rangle\right]\;,
\end{eqnarray}
and
\begin{eqnarray}\label{kubo-kap}
\kappa(\omega_c,T)=\frac{i}{\hbar\omega_c T \mathcal{L}}\left[ \langle
\hat\Theta_{xx}\rangle+\frac{\hbar}{\mathcal{Z}}
\sum_{n,m}\frac{e^{-\beta\varepsilon_n}-e^{-\beta\varepsilon_m}}
{\varepsilon_n-\varepsilon_m+\hbar\omega_c}|\langle n|\hat{J}^{Q}_x|m\rangle|^2\right]\;.
\end{eqnarray}
\end{widetext}
In the above, $|k\rangle$ is a normalized eigenstate of the
Hamiltonian with energy  $\varepsilon_k$, $\mathcal{Z}=\sum_k
\exp(-\beta\varepsilon_k)$  is the canonical partition function, 
and $\beta=1/k_BT$ is the inverse temperature.  A thermal 
average is indicated by $\langle\cdots\rangle$.  The dynamical temperature 
variation is turned on adiabatically from the infinite past, i.e., $\omega_c=\omega+i0^{+}$.

In Eqs.~\ref{kubo-sig}-~\ref{kubo-kap} the 
charge current $\hat{J}_x$ is formally given by
\begin{eqnarray}
\hat{J}_x=-\lim_{k_x\rightarrow0}\frac{d}{dk_x}[\hat K(k_x),q_e\hat{n}(-k_x)]
\label{j}
\end{eqnarray}
while the heat current $\hat{J}_x^Q$ is
\begin{eqnarray}
\hat{J}_x^Q=-\lim_{k_x\rightarrow0}\frac{1}{2}\frac{d}{dk_x}[\hat K(k_x),\hat
K(-k_x)]\;.
\end{eqnarray}
Here $\hat{K}=\hat{H} -\mu\hat{n}$ is the grand canonical Hamiltonian
and $\hat{A}(\vec k)=\sum\exp(i\vec k\cdot\vec
r)\hat{A}(\vec r)$ is the Fourier decomposition for mode $k$ of a local  
operator $\hat{A}(\vec r)$ which operates at position $\vec r$.  It should be noted that 
the heat and charge currents are related via the energy current as
$\hat{J}_x^Q=\hat{J}_x^E-(\mu/q_e)\hat{J}_x$ with
\begin{eqnarray}
\hat{J}_x^E=-\lim_{k_x\rightarrow0}\frac{d}{dk_x}[\hat T(k_x),\frac{1}{2}\hat
T(-k_x) + \hat U(-k_x)]\;,
\label{je}
\end{eqnarray}
with $\hat T$ and $\hat U$ being equal to 
the kinetic and potential energy operators\cite{T_U_note}, respectively.

From Ref.~\onlinecite{shastry_1,shastry_2,shastry_3} the definitions for
stress tensor $\hat{\tau}_{xx}$, the thermoelectric operator $\hat{\Phi}_{xx}$, 
and the thermal operator $\hat{\Theta}_{xx}$ are
\begin{eqnarray}
&&\hat{\tau}_{xx}=-\lim_{k_x\rightarrow0}\frac{d}{dk_x}[\hat{J}_x(k_x),q_e\hat n(-k_x)]\\
\label{phi}
&&\hat{\Phi}_{xx}=-\lim_{k_x\rightarrow0}\frac{d}{dk_x}[\hat{J}_x(k_x),\hat{K}(-k_x)]\\
&&\hat\Theta_{xx}=-\lim_{k_x\rightarrow0}\frac{d}{dk_x}[\hat{J}_x^Q(k_x),\hat{K}(-k_x)]\;.
\end{eqnarray}
The explicit form of these operators for the $t$-$J$ model are given in the appendix.

Throughout this work we take the chemical potential $\mu(T)$ in the above formulas 
to be that obtained from the canonical ensemble, i.e., $\mu(T)=\partial F/\partial N$, 
where $F=-(1/\beta)\ln(\mathcal{Z})$ 
is the Helmholtz free energy.  For a finite sized system we approximate this 
partial derivative as $\mu(T)=(F_{N+1}-F_{N-1})/2$ for an $N$ electron system~\cite{mu_note}.  

When calculating the full frequency dependent conductivities for finite sized clusters 
one must take into account the discreteness of the energy spectrum caused by the finite 
nature of the cluster.  This is done by introducing a broadening factor $\eta$ where 
the frequency then becomes $\omega_c\rightarrow\omega+i\eta$.  The broadening factor is 
taken to be the mean energy spacing between states with non-zero current 
matrix elements.  Table~\ref{eta_table} provides values of $\eta$ for the 
systems considered here and  $\eta$ is generally weakly 
dependent on $x$ and of order $\eta\sim3|t|$.

\begin{table}[ht]
\caption{Broadening factor $\eta$ for the $t$-$J$ model 
on a two-dimensional triangular lattice with $\mathcal{L}=12$ sites. Both 
positive and negative values of the hopping $t$ are given.  
The value of $J$ has very little effect on $\eta$ compared to the weak $x$ dependence.}
\begin{tabular}{|c|c|c|}
\hline
$x$&$\eta/|t|$ ($J=0.2|t|$,$t>0$)&$\eta/|t|$($J=0$,$t<0$)\\
\hline
0.83 & 3.229589 & 2.678842 \\
0.75 & 4.198734 & 4.221241 \\
0.67 & 4.659046 & 4.697231 \\
0.58 & 4.922330 & 4.934004 \\ 
0.17 & 3.786627 & 3.772738 \\   
0.083 & 2.803789 & 2.769104 \\   
\hline
\end{tabular}
\label{eta_table}
\end{table}

Following the work of Ref.~\onlinecite{shastry_1,shastry_2,shastry_3} one can
consider the high frequency  limit of the thermopower, Lorenz number,
and figure of merit in the hope and expectation that for strongly correlated systems 
modeled by the $t$-$J$ model these combinations of conductivities will have 
weak frequency dependence and yet still capture the essential 
strongly correlated physics, i.e., the Mott-Hubbard physics.  For notational convenience we 
indicate the high frequency expansions by a superscript star, that is, 
\begin{eqnarray}
S^*(T)=\lim_{w\rightarrow\infty}S(\omega,T)=
\frac{\langle\hat\Phi_{xx}\rangle}{T\langle\hat\tau_{xx}\rangle}\;,
\end{eqnarray}
\begin{eqnarray}
L^*(T)=\lim_{w\rightarrow\infty}L(\omega,T)=
\frac{\langle\hat\Theta_{xx}\rangle}{T^2\langle\hat\tau_{xx}\rangle}-\{S^*(T)\}^2\;,
\end{eqnarray}
and
\begin{eqnarray}\label{fom-hfe}
Z^*(T)T=\lim_{w\rightarrow\infty}Z(\omega,T)T=\frac{\{S^*(T)\}^2}{L^*(T)}\;.
\end{eqnarray}

While the high frequency starred quantities are not trivial to calculate 
they are considerably simpler than the full dynamical Kubo 
formulas (Eqs.~\ref{thermo}-\ref{fom}) as they 
are equilibrium expectation values and not dynamical in nature.  It is also 
reasonable to expect that in the future, approximate methods could be used 
to calculate their full temperature dependence that are not as 
limited as the exact diagonalization brute force methods used here.  However, one
must be very careful when aiming to establish new approximation for strongly correlated 
systems and, hence, the need for the brute force calculations in this work.

\section{Thermopower}\label{sec-s}

The thermopower can be factored instructively\cite{cw-prl} as
\begin{eqnarray}\label{s-factored}
S(\omega,T)=\frac{\gamma(\omega,T)}{\sigma(\omega,T)}=
\frac{\tilde\gamma(\omega,T)}{\sigma(\omega,T)}-\frac{\mu(T)}{q_eT}\;,
\end{eqnarray}
where $\gamma(\omega,T)=\tilde\gamma(\omega,T)-(\mu(T)/q_eT)\sigma(\omega,T)$ 
defines $\tilde\gamma(\omega,T)$.  This factorization 
displays clearly the two contributions composing the thermopower; one arising 
from electron transport and the other from thermodynamics (entropy).  This is discussed 
in detail below. 

As was mentioned previously, calculating this type of formula for strongly correlated systems 
is very difficult.  Often there is a desire to give the transport term little importance
and, therefore, drop it.  This leaves merely the second term in Eq.~\ref{s-factored} involving 
the chemical potential alone.  This is the so-called Mott-Heikes (MH) term for the 
thermopower which is valid at high temperatures and is described 
in more detail below.  However, as will be shown below, for low to intermediate temperatures 
this approximation is not adequate.

On physical, as well as theoretical\cite{stafford}, grounds, one 
expects the thermopower to vanish at $T=0$ and 
for a non-interacting system it is simple to show this.  For 
our purposes it is instructive to describe this vanishing through a delicate balancing act 
where the transport term exactly equals the MH term as the temperature tends towards zero, i.e., 
\begin{eqnarray}
\lim_{T\rightarrow0}\left\{\frac{\tilde\gamma(\omega,T)}{\sigma(\omega,T)}-
\frac{\mu(T)}{q_eT}\right\}=0\;.
\end{eqnarray} 

Achieving this balance in a finite sized system is not possible 
explicitly, even for non-interacting electrons.   
However, it does suggests a formulation of the thermopower into 
the two contributing terms mentioned above; a 
frequency dependent transport term $S_{tr}(\omega,T)$ and a frequency independent 
Mott-Heikes (MH) term $S_{MH}(T)$ with both defined through
\begin{eqnarray}
S(\omega,T)&=&\frac{1}{T}\left\{\frac{T\tilde\gamma(\omega,T)}{\sigma(\omega,T)}-
\frac{T\tilde\gamma(\omega,0)}{\sigma(\omega,0)}\right\}\nonumber\\
&&-\left\{\frac{\mu(T)-\mu(0)}{q_eT}\right\}\nonumber\\
&=&S_{tr}(\omega,T) + S_{MH}(T)\;.
\end{eqnarray}

Therefore, even for a finite sized system one can obtain a transport 
term and MH term that independently equal zero at $T=0$ ensuring that 
$S(\omega,T)$ vanishes in the zero temperature limit as expected.  This is not the complete 
picture however.  The MH term contains the chemical potential which is 
expected to behave quadratically in $T$ as $T\rightarrow0$ for thermodynamically 
large systems.  Finite systems have two particular differences.  One 
is that the spectrum is discrete giving rise to a ground state energy gap in situations 
without degeneracies.  In those instances there will be a low temperature 
exponential behavior of $\mu(T)$ which is not really 
a problem for our purposes because $\mu(T)/T$ will 
still vanish in the zero temperature limit.  The existence of ground state 
degeneracies, on the other hand, are a bigger concern.  Their existence cause the 
chemical potential to behave linearly in $T$ at low temperatures which, in turn, 
produces a MH term that does not vanish.  We argue that this is an 
unwanted unphysical result and our solution is to merely discount this ground 
state degeneracy (when it exists) when calculating $\mu(T)$ ensuring that the 
MH term vanishes as $T\rightarrow0$.

The high frequency expansion of $S(\omega,T)$ is similarly written as
\begin{eqnarray}
S^*(T)&=&\frac{1}{T}\left\{\frac{\langle\hat{\tilde\Phi}_{xx}(T)\rangle}{\langle\tau_{xx}(T)\rangle}-
\frac{\langle\hat{\tilde\Phi}_{xx}(0)\rangle}{\langle\tau_{xx}(0)\rangle}\right\}+S_{MH}(T)\nonumber\\
&=&S^*_{tr}(T)+S_{MH}(T)\;.
\end{eqnarray}
Again we have defined $\hat{\tilde{\Phi}}_{xx}$ similarly to $\tilde\gamma(\omega,T)$ 
through $\hat\Phi_{xx}=\hat{\tilde\Phi}_{xx}-(\mu(T)/q_e)\hat\tau_{xx}$.

The transport term of the thermopower eventually vanishes as $T$ becomes 
large so we know that the Mott-Heikes term eventually dominates 
the thermopower and becomes useful for a number of reasons. One reason it is that 
it is not a dynamical 
quantity and, hence, is easier to compute.  Secondly, it is often hoped 
that the MH term dominates the thermopower and one only needs 
to consider it.  This is due to the fact that at high temperatures it
approaches a constant since $\mu(T)$ is eventually linear in $T$.  Previous 
work by Beni\cite{beni}, and Chaikin and Beni\cite{beni_chaikin} worked out the 
infinite temperature limit of $S_{MH}(T)$ for a number of systems.  There is an 
elegance and simplicity to these formulas since the infinite temperature limit 
of $S_{MH}(T)$ is determined merely from counting arguments related to the Hilbert 
space dimension of the problem at hand.  A central question regarding the MH term is 
how low of a temperature does the MH limit remain a valid approximation to 
the full thermopower.  We provide an answer to that question 
for the $t$-$J$ model in this work which is discussed later.  

The two MH limits we consider 
here are for the uncorrelated band\cite{mh_note} and 
for the $t$-$J$ model\cite{shastry_1,beni_chaikin,mukerjee},
\begin{widetext}
\begin{eqnarray}
\lim_{T\rightarrow\infty}S_{MH}(T) = \left\{ \begin{array}{l l}
  \frac{k_B}{q_e}\ln\left(\frac{2-n}{n}\right) & \quad \mbox{uncorrelated with $0\leq n\leq2$}\\
  \frac{k_B}{q_e}\ln\left(\frac{2(1-n)}{n}\right) & \quad \mbox{$t$-$J$ with $0\leq n\leq1$} \\
  -\frac{k_B}{q_e}\ln\left(\frac{2(n-1)}{2-n}\right) & \quad \mbox{$t$-$J$ with $1\leq n\leq2$}\\
\end{array}
\label{mh-eq}
\right.
\end{eqnarray}
\end{widetext}
remembering that $q_e=-|e|$ is the electric charge.

Although a somewhat blunt formulation, the Mott-Heikes limits already contain a 
plethora of information.  For example, for the uncorrelated model, 
even with a finite interaction parameter $U$, the thermopower 
diverges at $x\rightarrow 1$ ($n\rightarrow0$), is positive for all hole dopings $x>0$ ($n<1$) 
and is exactly zero for the half filled case $x=0$ ($n=1$).  
For electron doping the thermopower would be purely negative 
diverging negatively as $x\rightarrow1$ ($n\rightarrow2$).  In the whole range of densities ($0\leq n\leq2$) the 
MH limit would predict a single sign change.  

For the $t$-$J$ model the essentially infinite 
strength interactions of the electrons causes two additional sign changes compared 
to the uncorrelated or finite $U$ Hubbard model.  The thermopower still 
diverges as $x\rightarrow1$, is positive for $x>1/3$, 
negative for $x<1/3$, and diverges negatively at half filling.  For electron doping we use particle 
hole symmetry to get precisely the opposite behavior: a positive divergence at half filling, positive for 
$x<1/3$, a sign change to negative thermopower for $x>1/3$ and a negative divergence as 
$x\rightarrow1$.  Hence, two additional zero crossings emerge due to the interactions.

In the following subsections we report results for 
$S^*(T)$, the difference $S(\omega,T)-S^*(T)$ and the Mott-Heikes term $S_{MH}(T)$ 
for both positive (Sec.~\ref{s_pos_t}) and negative (Sec.~\ref{s_neg_t}) 
signs of the hopping $t$.  In all figures the thermopower is given in 
experimental units of $\mu V/K$ where $k_B/|q_e|=86\mu V/K$, and we have 
multiplied by ($-1$) to facilitate comparison with the electron doped 
NCO system, see Eq.~\ref{mh-eq}.

\subsubsection{Positive hopping $t>0$}\label{s_pos_t}

\begin{figure}[t]
\begin{center}
\mbox{\bf (a)} 
\includegraphics[width=8.cm]{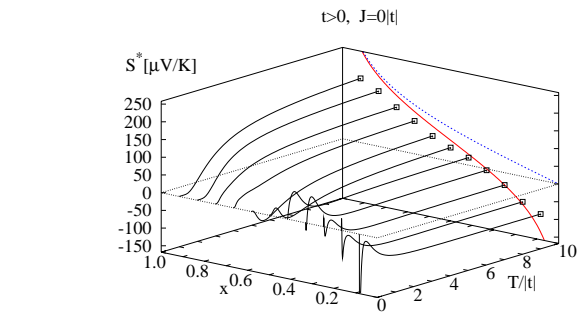}\\
\mbox{\bf (b)} 
\includegraphics[width=8.cm]{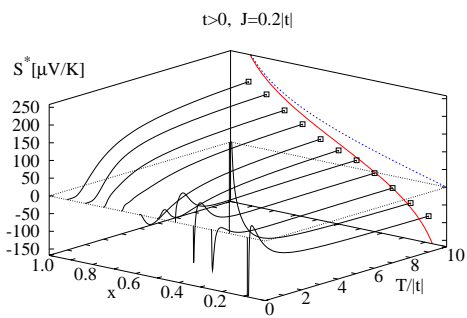}
\end{center}
\caption{(color online) $S^*(T)$ as a function of doping $x$ and temperature $T$ 
for positive hopping $t>0$ corresponding to NCO after 
particle-hole transformation.  Panel {\bf (a)} and {\bf (b)} are for $J=0$ and 
$0.2|t|$, respectively.  Projected onto the $T=10|t|$ plane is 
the Mott-Heikes limits for the uncorrelated (blue dotted) and the $t$-$J$ 
models (red), respectively.  $S^*(T)$ approaches 
the MH limit for the $t$-$J$ model relatively quickly, i.e., by 
approximately $T\sim6|t|$.  The horizontal black dotted lines indicate 
the position of zero thermopower.} 
\label{sstar_t1}
\end{figure}

\begin{figure}[t]
\begin{center}
\mbox{\bf (a)} 
\includegraphics[width=8.cm]{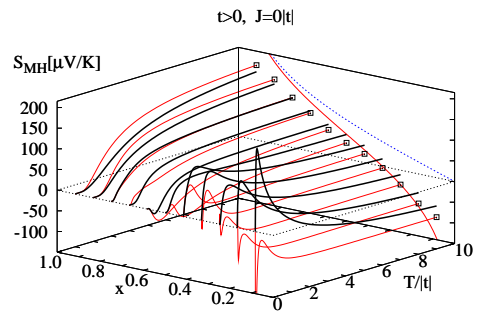}\\
\mbox{\bf (b)} 
\includegraphics[width=8.cm]{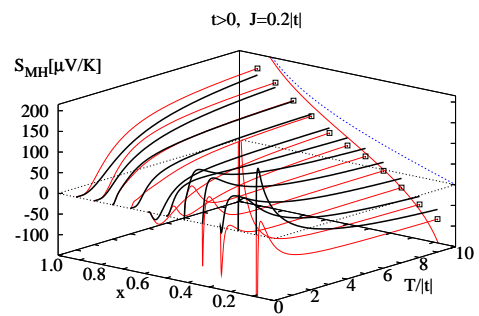}
\end{center}
\caption{(color online) The Mott-Heikes term of the thermopower $S_{MH}(T)$ versus 
doping $x$ and temperature $T$ for positive hopping $t>0$ corresponding to NCO after 
particle-hole transformation.  Panel {\bf (a)} and {\bf (b)} are for $J=0$ 
and $0.2|t|$, respectively.  The black line is $S_{MH}(T)$ while the 
red line is the full $S^*(T)$.  The MH limits are projected onto the $T=10|t|$ plane.}
\label{sstar_mh_t1}
\end{figure}

Fig.~\ref{sstar_t1} shows $S^*(T)$ as a function of both doping and temperature for 
the case of positive hopping $t>0$.  We have 
computed $S^*(T)$ for two different values of $J$, namely, $J=0$ (Fig.~\ref{sstar_t1}a) and 
$J=0.2|t|$ (Fig.~\ref{sstar_t1}b).  Projected onto the $T=10|t|$ plane are the 
two MH limits; the uncorrelated model (blue dashed line) and the $t$-$J$ 
model (solid red line).  The high temperature behavior of $S^*(T)$ matches the 
$t$-$J$ model MH limit expectation quite satisfactorily producing 
a sign change near $x=1/3$.  The slight difference between $S^*(T=10|t|)$ is 
a combination of $T=10|t|$ being large but finite and 
finite size effects coming from the finite dimension of the Hilbert space 
yielding a discrepency even at $T=\infty$.

For large dopings $x\ge0.58$ the thermopower monotonically grows to a somewhat large 
value of $(100-200)\mu V/K$, growing faster with temperature the higher the 
doping.  The value of $J$ has little to no effect in this range of doping.  The thermopower is 
pinned at zero for $T=0$ and needs to eventually increase to its MH limit which in this range of doping is 
positive and large.  Since the doping is large there is little interaction 
between electrons and evidently they effectively avoid one another.  Hence the 
transport term $S^*_{tr}(T)$ has very little impact on the full thermopower.  
This physics is borne out by comparing $S^*(T)$ to only the Mott-Heikes term $S_{MH}(T)$ which is 
shown in Fig.~\ref{sstar_mh_t1}a-b where there is very little difference between 
$S^*(T)$ and $S_{MH}(T)$.

\begin{figure*}[t]
\centering
\begin{tabular}{cc}
\mbox{\bf (a)} 
\includegraphics[width=7.5cm]{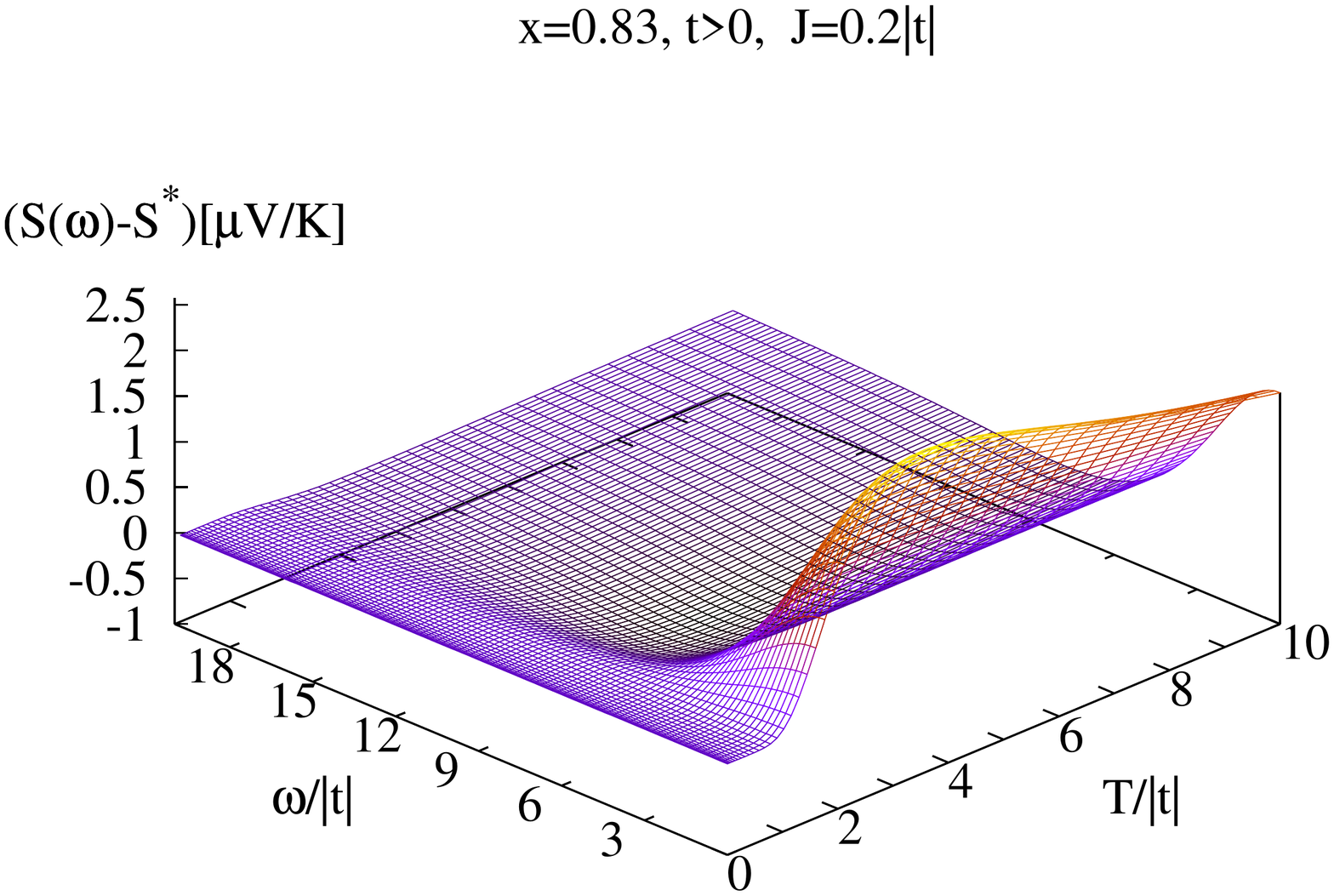}&
\mbox{\bf (b)} 
\includegraphics[width=7.5cm]{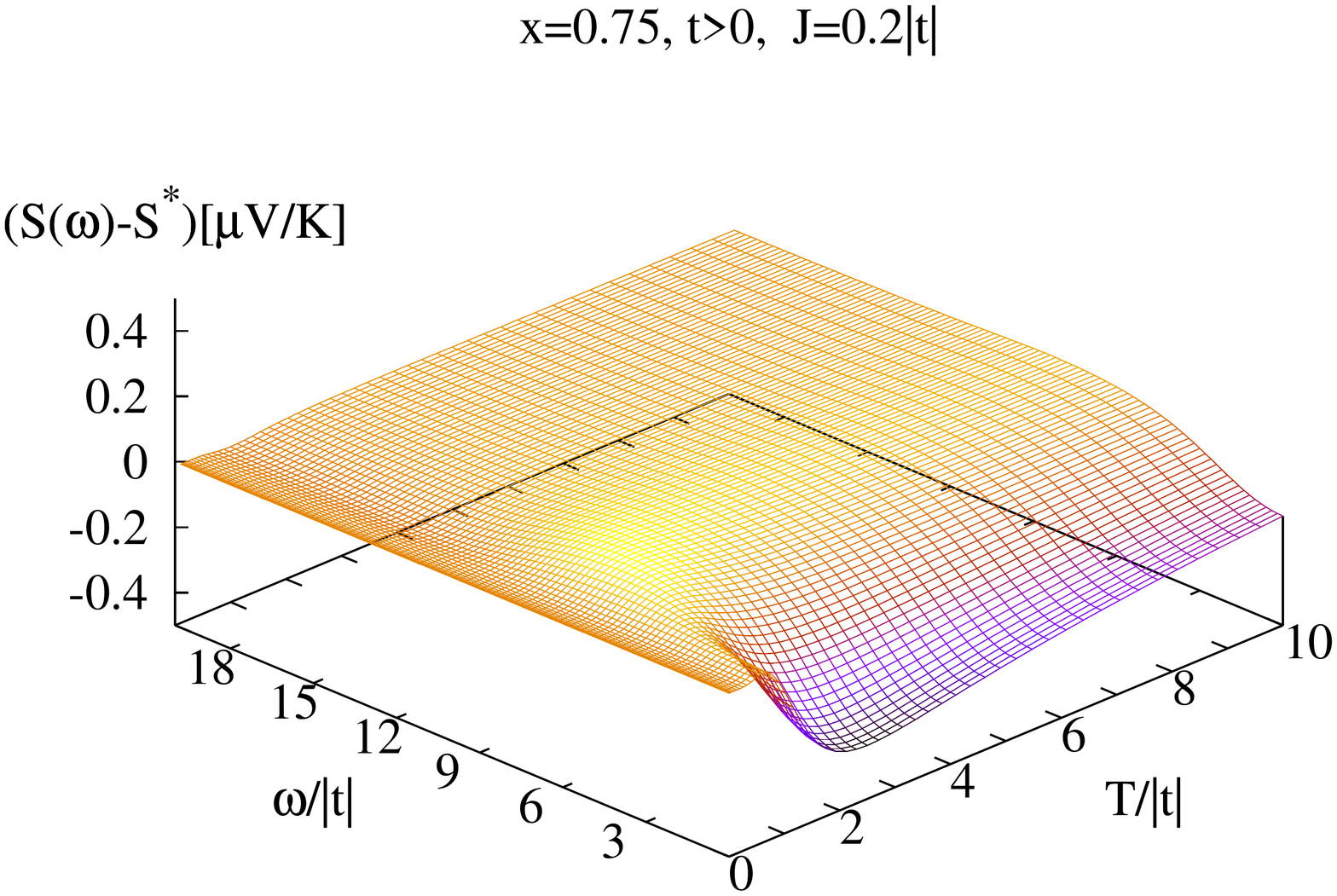}\\
\mbox{\bf (c)} 
\includegraphics[width=7.5cm]{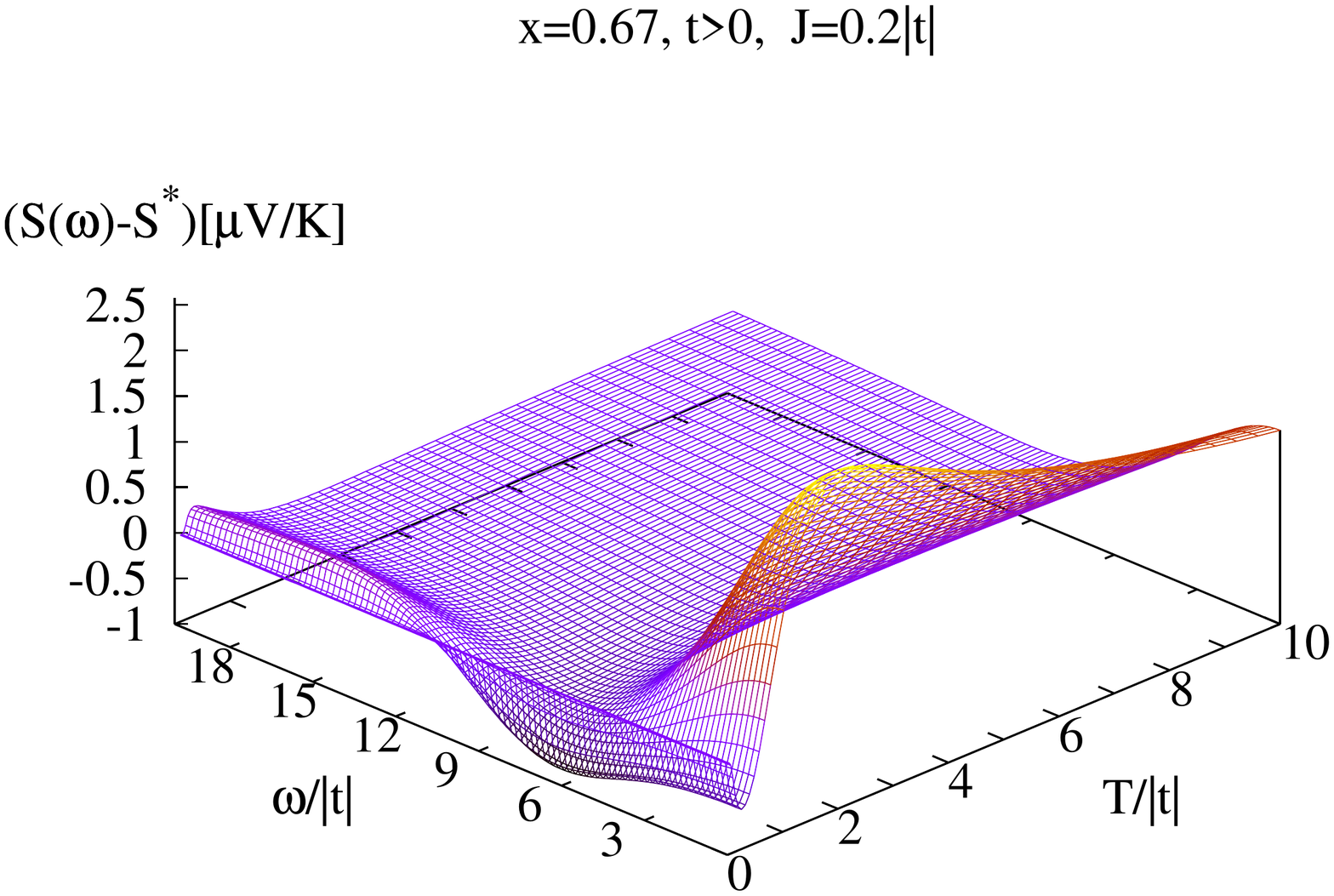}&
\mbox{\bf (d)} 
\includegraphics[width=7.5cm]{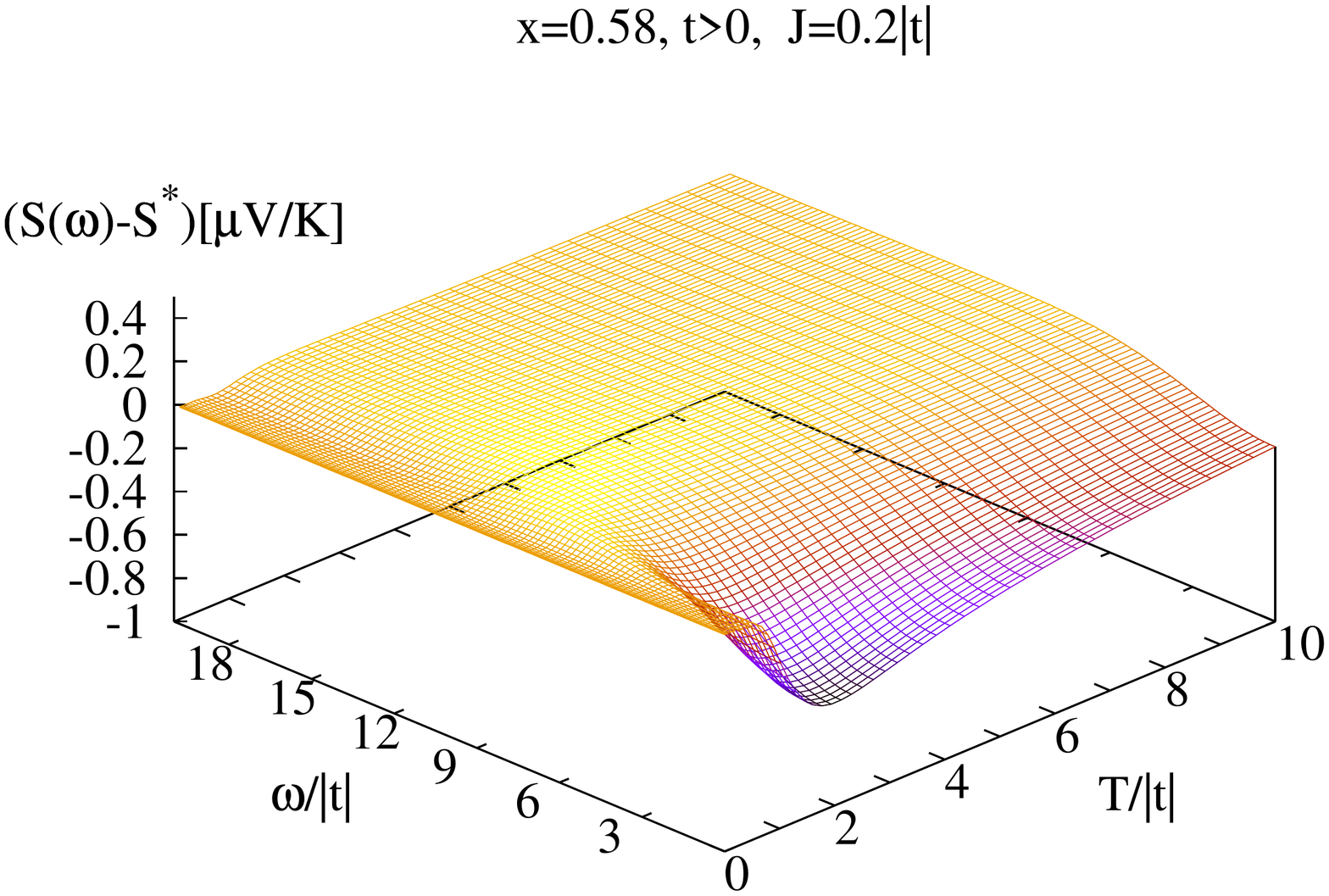}\\
\mbox{\bf (e)} 
\includegraphics[width=7.5cm]{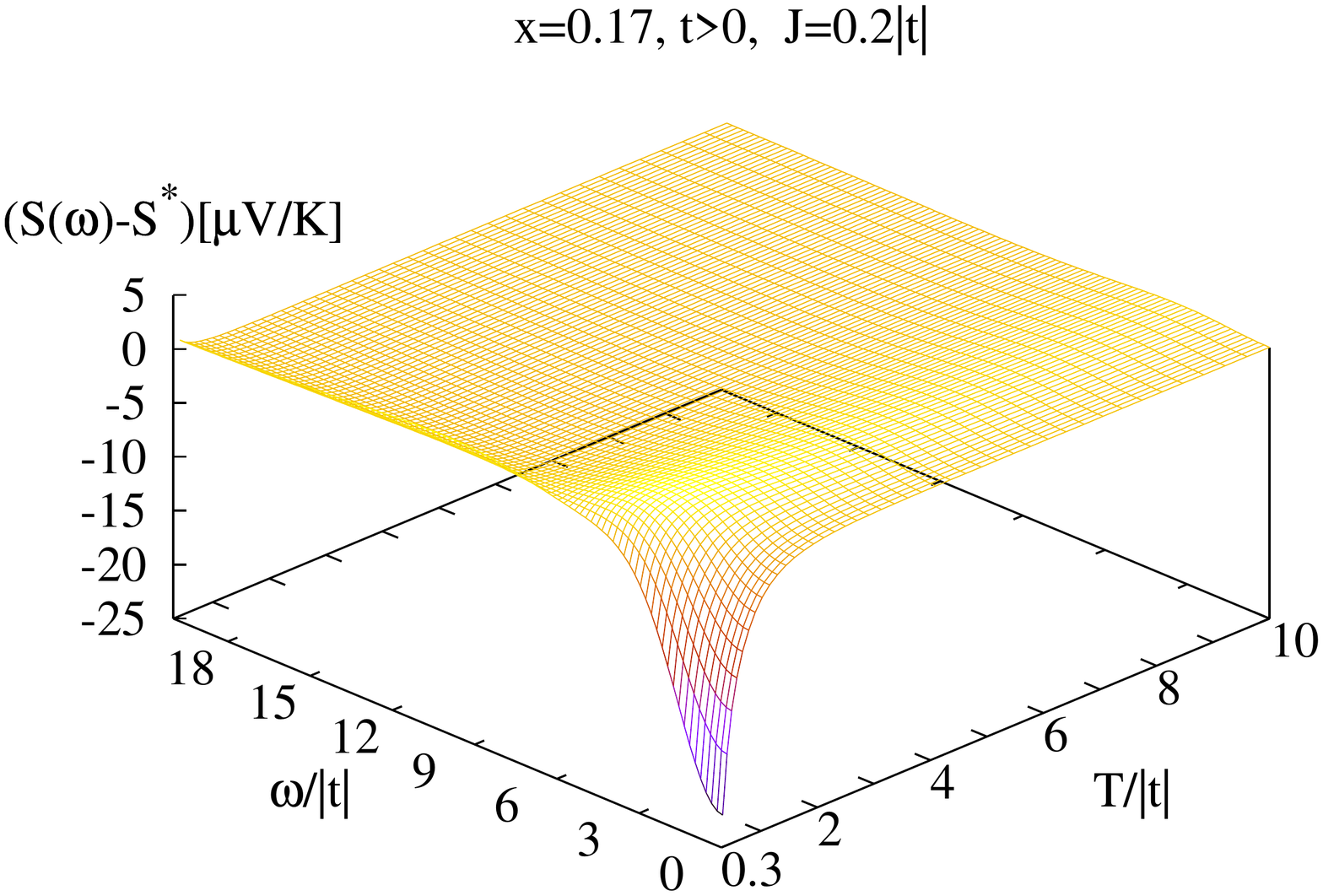}&
\mbox{\bf (f)} 
\includegraphics[width=7.5cm]{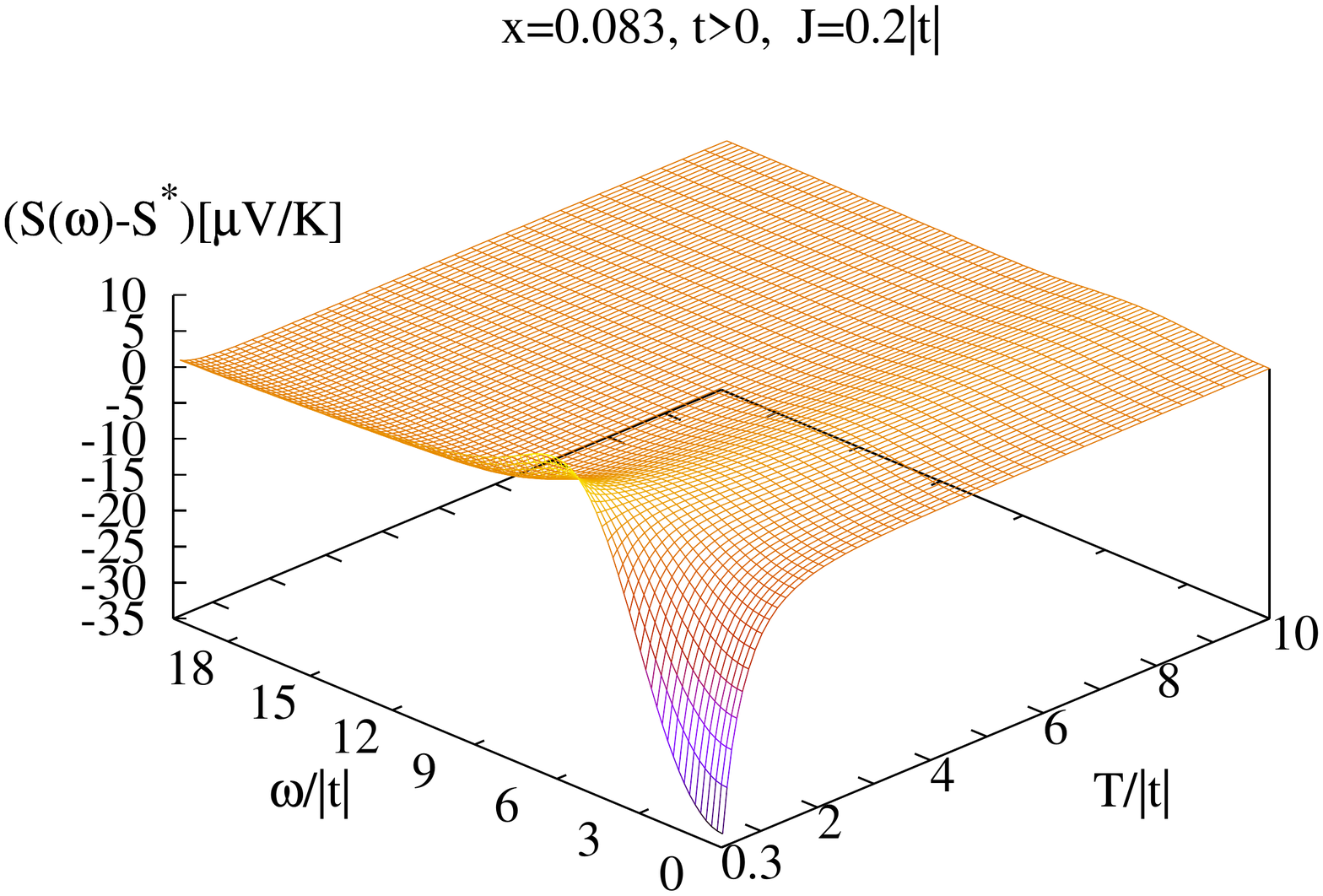}
\end{tabular}
\caption{(color online) $S(\omega,T)-S^*(T)$ as a function of frequency $\omega$ and temperature $T$ 
for $J=0.2|t|$, positive sign of the hopping $t>0$ (corresponding to NCO after 
particle-hole transformation) and for dopings {\bf (a)} $x=0.83$, {\bf (b)} $x=0.75$, {\bf (c)} $x=0.67$, 
{\bf (d)} $x=0.58$, {\bf (e)} $x=0.17$, and {\bf (f)} $x=0.083$.  The frequency 
dependence is evidently quite weak for 
dopings $x\geq0.58$ ({\bf (a)}-{\bf (d)}).  For $x=$0.17, and 0.083 ({\bf (e)} and {\bf (f)}) 
there is a much stronger frequency dependence that occurs at 
extremely small $\omega$ and $T$.  This is most 
likely due to finite size effects of our $\mathcal{L}=12$ site lattice and 
not an intrinsic property of the $t$-$J$ model at these dopings.  For 
parameters $\omega>3|t|$ and $T>2|t|$ the frequency dependence is approximately 
flat.  The doping of $x=0.92$ corresponds in our case to only one electron 
and, hence, has no frequency dependence.}
\label{s-sstar_t1}
\end{figure*}

For dopings $x\leq0.5$ strong electron correlation effects
are obtained.  This is due to the transport term which acts to 
reduce the thermopower at low to intermediate temperatures.  $S^*(T)$  
no longer monotonically approaches its MH limit.  Again this is shown more 
distinctly when 
one compares $S^*(T)$ to $S_{MH}(T)$ in Fig.~\ref{sstar_mh_t1}a-b where 
the MH term overestimates the thermopower indicative of 
a very active and important transport term $S^*_{tr}(T)$ which serves to reduce 
the thermopower.  Eventually, the transport term vanishes as $T$ becomes large and the MH term again 
dominates.

As the doping approaches half filling ($x\rightarrow 0$)
the thermopower begins to be purely negative and nearly monotonically 
approaches its now negative MH limit.  Some of the violent behavior at the lowest 
temperatures reported is no doubt due to peculiarities of the finite 
sized lattice.  That aside, the transport term has an increasingly important role to 
play at low to intermediate temperatures as the doping is reduced.  Again, this is quite obvious 
in Fig.~\ref{sstar_mh_t1}a-b where the MH and transport terms are quite divergent.  
Hence, it is definitely not a good approximation 
to use $S_{MH}(T)$ as a representative of the thermopower in low 
doping regions of strongly correlated systems for low 
to intermediate temperatures ($T\leq 5|t|$). 

The value of $J$ has almost no effect on the the thermopower until 
a doping of $x=0.25$ is reached.  Not shown are similar results 
for $J=0.4|t|$.

It is interesting to note that $S^*(T)$ can be well approximated 
by the infinite temperature MH limit for all dopings and all $J$ 
when the temperature is at or above approximately $5|t|\lesssim T \lesssim 6|t|$, as the thermopower 
has an overwhelming MH constribution by that temperature.

\begin{figure}[t]
\begin{center}
\mbox{\bf (a)} 
\includegraphics[width=8.cm]{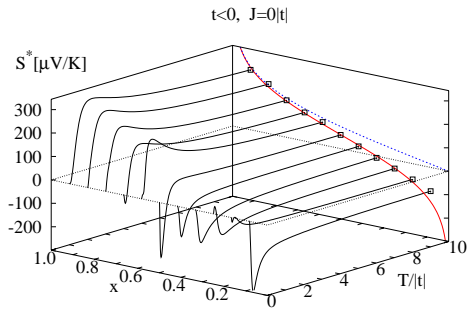}\\
\mbox{\bf (b)} 
\includegraphics[width=8.cm]{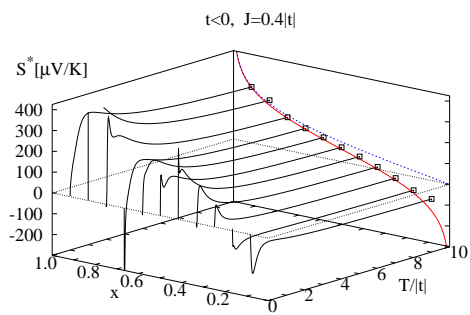}
\end{center}
\caption{(color online) $S^*(T)$ as a function of doping $x$ and temperature $T$ 
for negative hopping $t<0$ corresponding to a fiduciary hole 
doped CoO$_2$ compound.  Panel {\bf (a)} and {\bf (b)} are for $J=0$ and 
$0.4|t|$, respectively.  Projected onto the $T=10|t|$ plane is 
the MH limits.  $S^*(T)$ approaches the MH limit for the $t$-$J$ model relatively quickly, i.e., by 
approximately $T\sim6|t|$.  The horizontal black dotted lines indicate 
the position of zero thermopower.} 
\label{sstar_t-1}
\end{figure}

\begin{figure}[t]
\begin{center}
\mbox{\bf (a)} 
\includegraphics[width=8.cm]{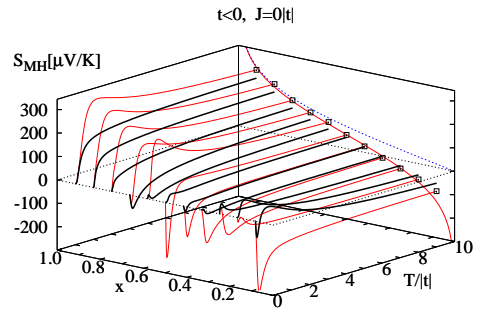}\\
\mbox{\bf (b)} 
\includegraphics[width=8.cm]{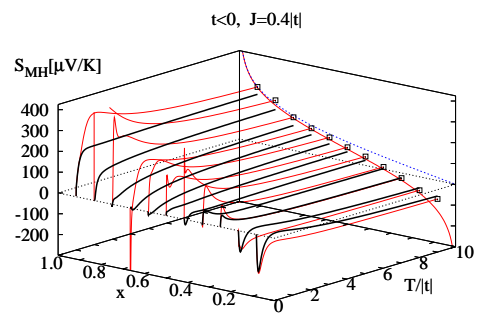}
\end{center}
\caption{(color online) The Mott-Heikes term of the thermopower $S_{MH}(T)$ versus 
doping $x$ and temperature $T$ for negative hopping $t<0$ corresponding to a fiduciary hole 
doped CoO$_2$ compound.  Panel {\bf (a)} and {\bf (b)} are for $J=0$ 
and $0.4|t|$, respectively.  The black line is $S_{MH}(T)$ while the 
red line is the full $S^*(T)$.  The MH limits are projected onto the $T=10|t|$ plane.}
\label{sstar_mh_t-1}
\end{figure}

Lastly we justify the use of $S^*(T)$ instead of the full dynamical 
thermopower $S(\omega,T)$.  Fig.~\ref{s-sstar_t1}a-f shows 
$S(\omega,T)-S^*(T)$ as a function of temperature and 
frequency $\omega$ for dopings $x=$0.83, 0.75, 0.67, 0.58, 0.17, and 
0.08.  The dopings $x=$0.5, 0.42, 0.33, and 0.25 cannot 
be calculated at this time due to computational constraints, i.e., the double 
sum over the current matrix elements in Eqs.~\ref{kubo-sig} and \ref{kubo-gam} is 
quite prohibitive.  For $x\leq 0.58$ the difference between the dynamical thermopower and the 
infinite frequency expansion is less than $2.5\mu V/K$ and hence 
has very little absolute effect (approximately less 
than $\sim 2\%$ difference).  Further it should be noted that for values 
of $\omega>3|t|$ and temperatures $T>2|t|$ the frequency dependence of 
the thermopower is nearly nonexistent.  

For smaller dopings, i.e., $x=$0.17, and 0.083, a more severe difference 
between $S(\omega,T)$ and $S^*(T)$ is found.  However, this larger difference takes place 
at extremely small frequencies and temperatures especially considering this calculation 
is done for a finite sized system.  Recall from Fig.~\ref{sstar_t1}b that at temperatures 
below approximately $T\sim0.3|t|$ the thermopower displays a drastic behavior 
as it approaches $T=0$.  This extreme behavior is almost certainly a consequence of the 
finite sized lattice on which we work and not an intrinsic property 
of the $t$-$J$ model.  In fact, in Fig.~\ref{s-sstar_t1}e and~\ref{s-sstar_t1}f we have 
cut the temperature off below $T=0.3|t|$ as the thermopower is badly divergent.  
Therefore, it should be concluded that the frequency dependence 
of $S(\omega,T)$ is most likely weak even for very small dopings.

\subsubsection{Negative hopping $t<0$}\label{s_neg_t}

\begin{figure*}[t]
\centering
\begin{tabular}{cc}
\mbox{\bf (a)} 
\includegraphics[width=7.5cm]{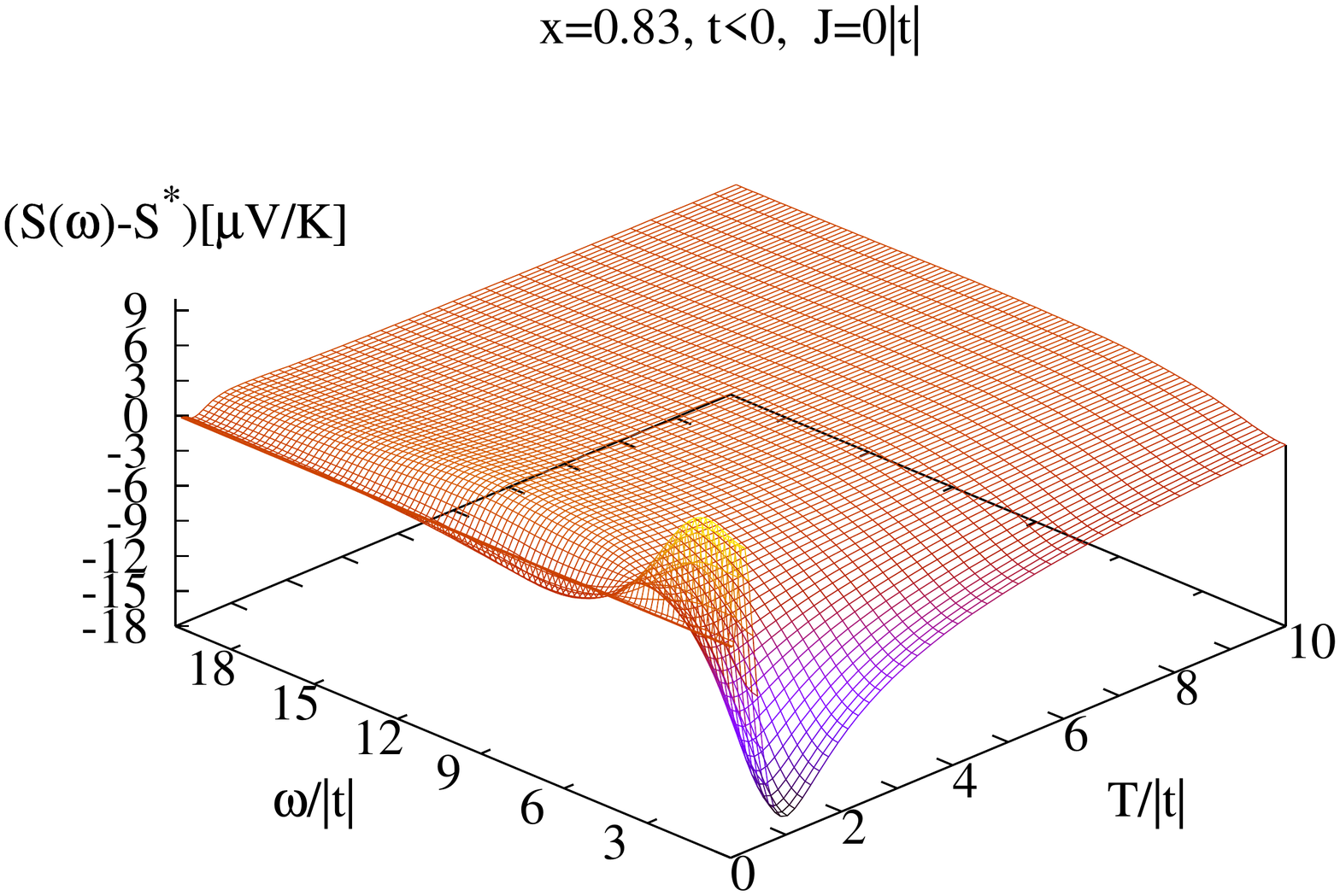}&
\mbox{\bf (b)} 
\includegraphics[width=7.5cm]{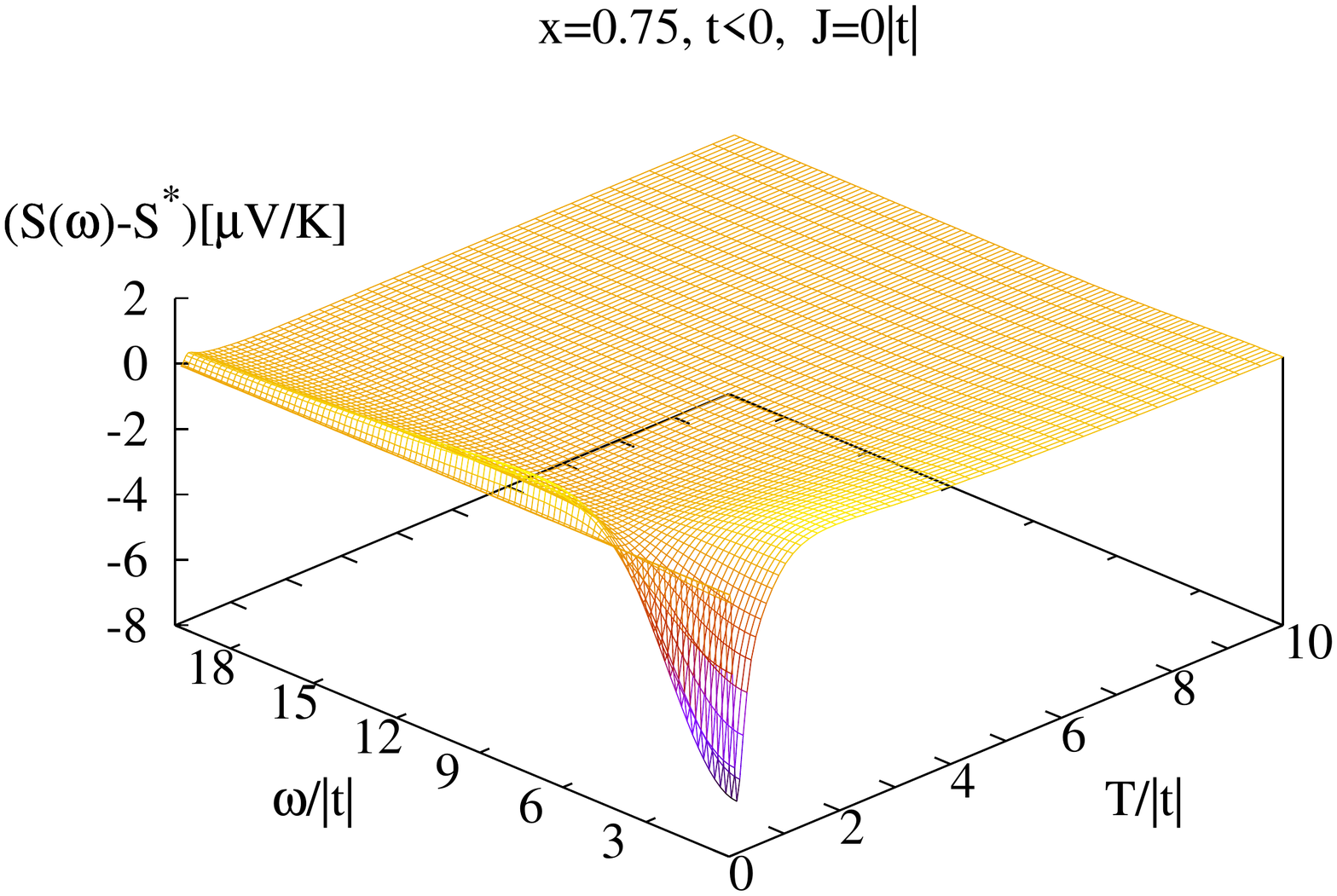}\\
\mbox{\bf (c)} 
\includegraphics[width=7.5cm]{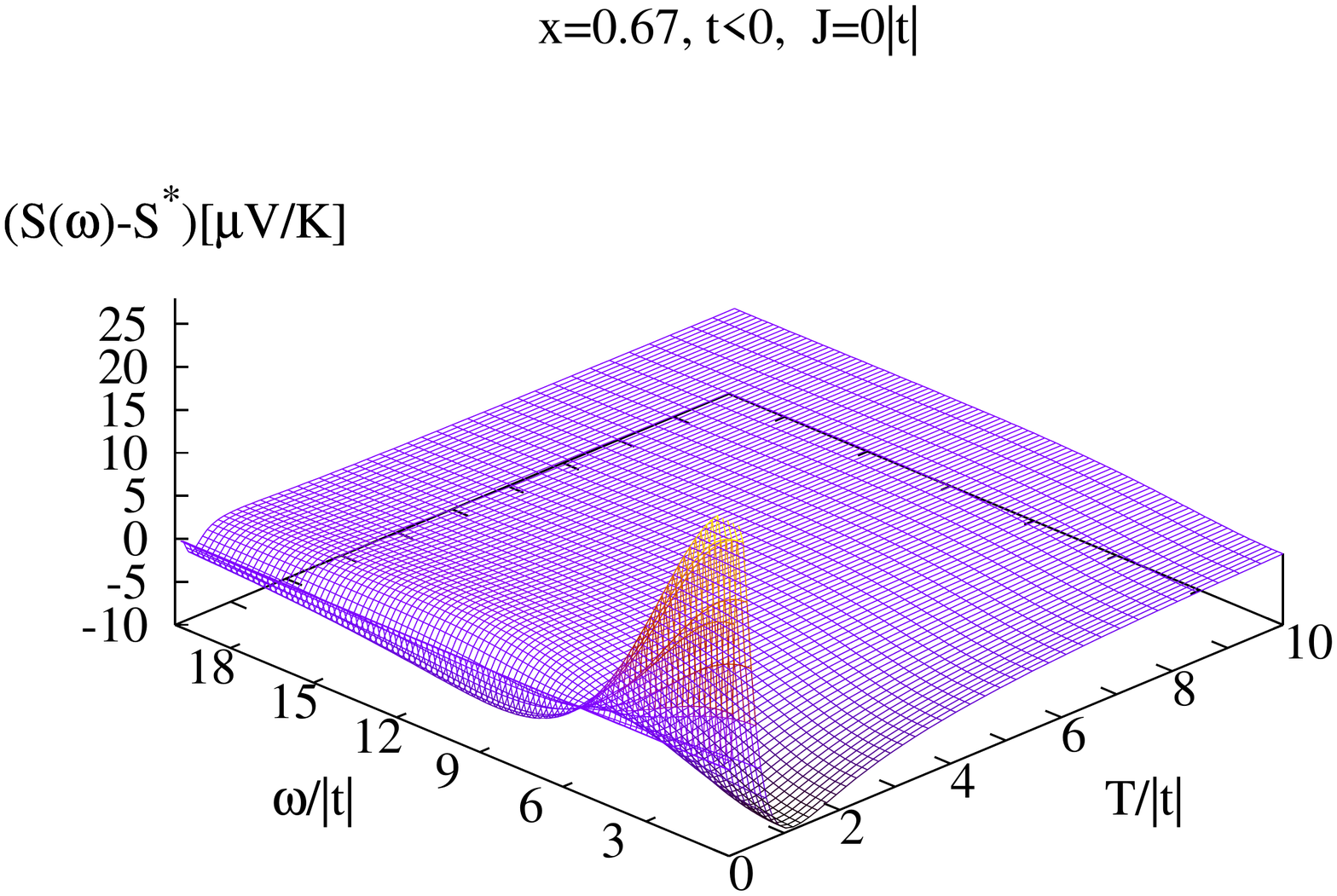}&
\mbox{\bf (d)} 
\includegraphics[width=7.5cm]{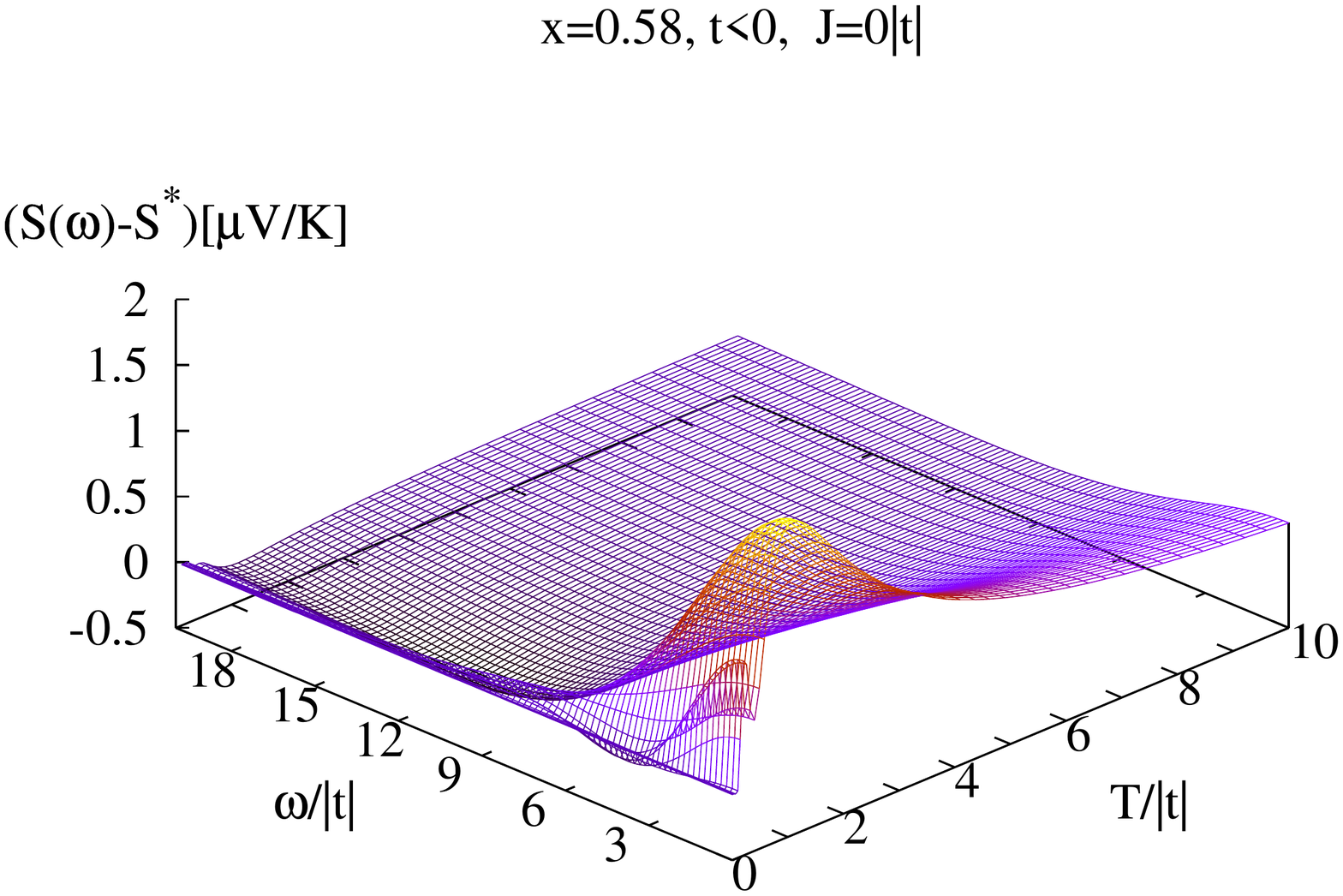}\\
\mbox{\bf (e)} 
\includegraphics[width=7.5cm]{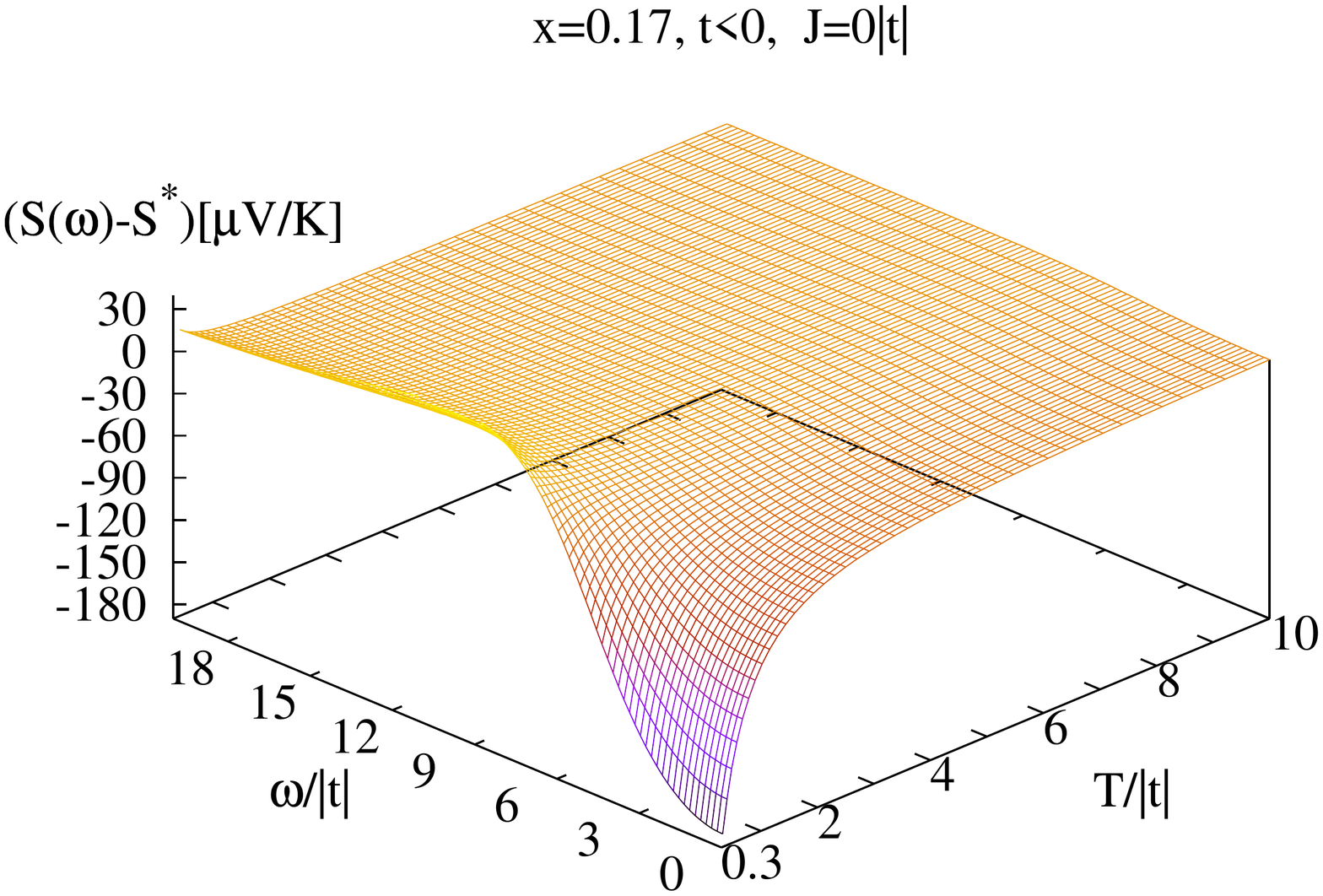}&
\mbox{\bf (f)} 
\includegraphics[width=7.5cm]{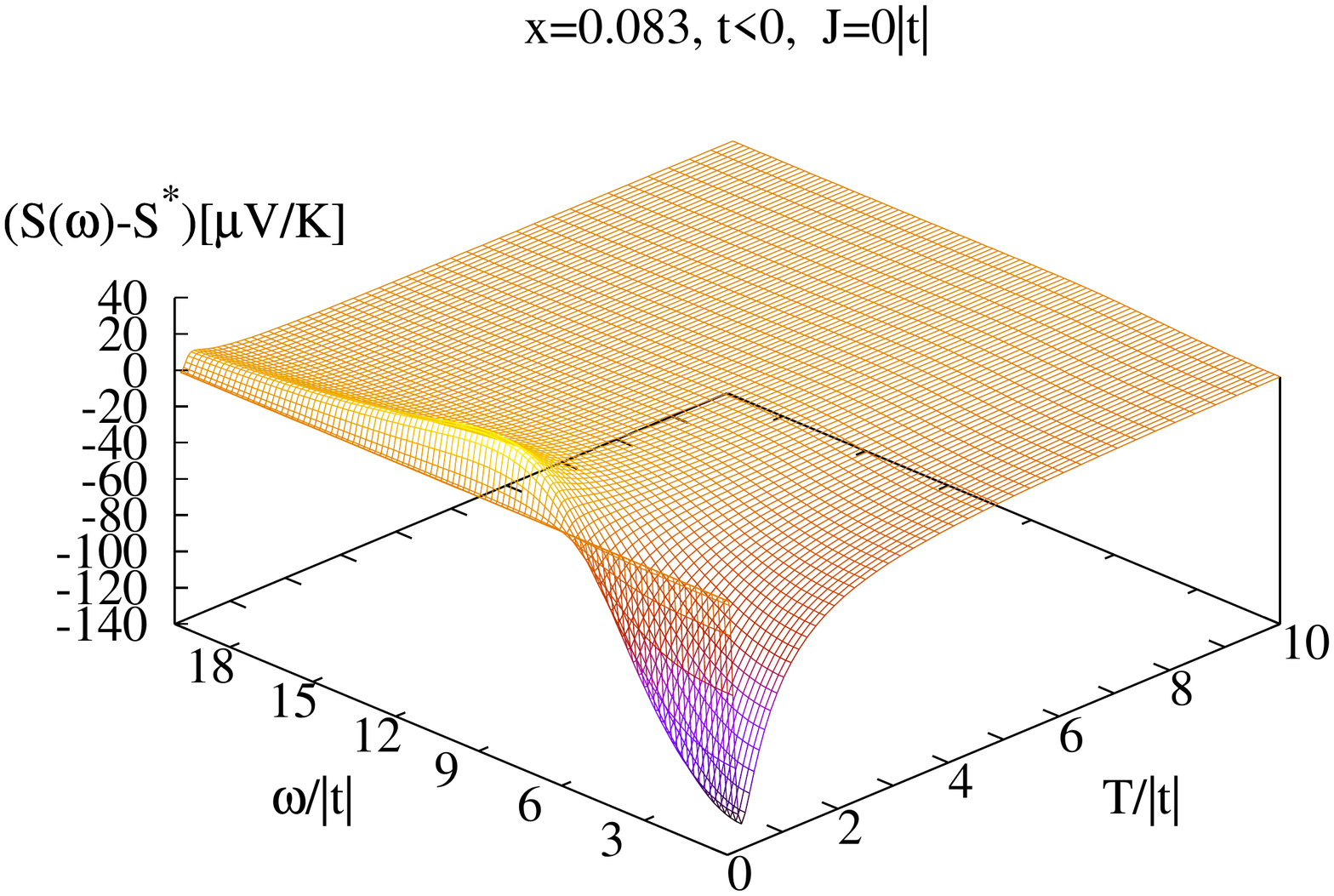}
\end{tabular}
\caption{(color online) $S(\omega,T)-S^*(T)$ as a function of frequency $\omega$ and temperature $T$
for $J=0$, negative sign on the hopping $t<0$ (corresponding to a fiduciary hole 
doped CoO$_2$ compound) and dopings {\bf (a)} $x=0.83$, {\bf (b)} $x=0.75$, {\bf (c)} $x=0.67$, 
{\bf (d)} $x=0.58$, {\bf (e)} $x=0.17$, and {\bf (f)} $x=0.083$.  The general frequency dependence is 
quite similar, albeit larger (discussed further in the text) to that 
shown in Fig.~\ref{s-sstar_t1} for positive hopping.}
\label{s-sstar_t-1}
\end{figure*}

We now consider the thermopower for a fiduciary system where we have switched the 
sign of the hopping $t$, i.e., $t<0$.  Since we study $0\leq n\leq1$, by a particle-hole transformation, 
this corresponds to $t>0$ for $1\leq n\leq2$, which differs from Na$_x$CoO$_2$ in the sign of the 
hopping ($t<0$ for NCO).  Such a system does not exist in the laboratories, and our hope is that 
our result will stimulate the search for a hole doped CoO$_2$ system.

In view of the topology of the geometrically 
frustrated triangular lattice there is the possibility of important effects coming 
from primarily the transport term.  In Ref~\onlinecite{shastry_1}, Shastry 
obtained a high temperature expansion of $S^*(T)$ for the triangular lattice 
which we quote here for completeness for the situation corresponding to $0\leq n\leq 1$ 
and $t>0$,
\begin{eqnarray}\label{hte}
S^*(T)&=&\frac{k_B}{q_e}\left\{\ln\left(\frac{2x}{1-x}\right)-
t\frac{1+x}{2T}\right\}\nonumber\\
&&+\mathcal{O}(\beta^2t^2)\;,
\end{eqnarray}
with hole doping $x=1-n$.  Comparision with an electron doped system such as 
NCO is facilitated through the particle-hole transformation given 
previously, c.f. Sec.~\ref{sec-model}.  The first term in Eq.~\ref{hte} is the 
Mott-Heikes term.  The second term is 
due to the transport and serves to reduce the thermopower from its large MH 
upper limit as the temperature is decreased.  Importantly, it depends on a single power 
of the hopping $t$.  Hence, if one could switch the sign of the hopping 
then $S^*(T)$ would evidently grow to a maximum as the temperature was decreased before 
approaching its zero value at $T=0$.  A term similar to the second term in  Eq.~\ref{hte} with 
an odd power of $t$ is only present in cases where the topology of the underlying lattice is 
geometrically frustrated.

Fig.~\ref{sstar_t-1}a-b shows $S^*(T)$ for the case of negative hopping as a 
function of $x$ and $T$ for $J=0$ and $J=0.4|t|$, respectively (recall 
that we actually plot $(-1)\times S^*(T)$ to facilitate comparison 
with NCO).  The prediction of 
thermopower enhancement is quite clearly visible.  For dopings $x>0.5$, 
instead of monotonically decreasing from its upper MH limit the thermopower 
grows to a maximum at approximately $1|t|\lesssim T\lesssim 2|t|$ before being pinned by 
its $T=0$ constraint.  The enhancement is also larger for larger dopings.  Again 
the value of $J$ has little effect other than at extremely 
low temperatures which most assuredly suffer from finite size effects.  As expected 
the thermopower obtains its MH limit when $T$ reaches $5|t|\lesssim T\lesssim 6|t|$ just as 
for the positive hopping situation.

The origin of the thermopower enhancement stems from the 
transport term $S^*_{tr}(T)$ as expected.  Fig.~\ref{sstar_mh_t-1}a-b 
shows the Mott-Heikes formula for the thermopower along with the 
full $S^*(T)$ for the negative hopping case.  The MH term clearly underestimates the magnitude of the 
enhancement and nearly misses it all together.  Obviously the transport 
term has a much more sensitive dependence on the lattice topology than the 
MH term.

The magnitude of the enhancement is also quite striking.  A value of thermopower 
greater than 150 $\mu V/K$ is already anomalously large for a 
seemingly metallic system such as this.  However, upon switching the 
sign of the hopping parameter a value of the thermopower of nearly 
350 $\mu V/K$ is obtained.  Clearly, one should consider the 
intermediate temperature transport effects when attempting to discover/design
large thermopower materials; for NCO ($|t|\sim 100K$\cite{shastry_1,shastry_2,cw-prl,cw-prb}) the 
intermediate temperature range corresponds to approximately room temperature.

The intermediate temperature enhancement of the thermopower is 
very similar in magnitude and shape to the enhanced thermopower of 
NCO at high dopings observed recently by Lee, et. al.\cite{minhyea-lee}.  Although 
there is, at present, no reason to believe that the experimental system 
has an inversion of the sign of the hopping as the doping is increased from 
the Curie-Weiss metallic phase, the similarity between our calculation 
and the data is striking.

For completeness, Fig.~\ref{s-sstar_t-1}(a-f) displays 
the full frequency dependence of $S(\omega,T)$ compared to $S^*(T)$ for the 
negative hopping.  The same general behavior is shown compared to the positive 
hopping situation in Fig.~\ref{s-sstar_t1}(a-f).  The main difference 
between the two cases is that the deviation of $S(\omega,T)$ and $S^*(T)$ is slightly 
more extreme.  However, similarly, this is seen to occur at low temperatures, frequencies, 
and dopings, and is most likely a finite size artifact.

\section{Lorenz number}\label{sec-l}

The Lorenz number is an important thermoelectric quantity, not least of 
which, is its importance in the dimensionless 
figure of merit (FOM).  The FOM is an extremely important quantity when 
determining the technological usefulness and performance of 
a thermoelectric material where values of FOM in excess of unity are 
highly desirable.

Recall the definition of the Lorenz number from Eq.~\ref{lorenz}
\begin{eqnarray}
L(\omega,T)=\frac{\kappa(\omega,T)}{T\sigma(\omega,T)}-S(\omega,T)^2\;,
\end{eqnarray}
which can be simplified as
\begin{eqnarray}
L(\omega,T)=\frac{\tilde\kappa(\omega,T)}{T\sigma(\omega,T)}-
\left\{\frac{\tilde\gamma(\omega,T)}{\sigma(\omega,T)}\right\}^2
\end{eqnarray}
with $\kappa(\omega,T)=\tilde\kappa(\omega,T)-(2\mu(T)/q_e)\tilde\gamma(\omega,T)
+(\mu(T)/\sqrt{T}q_e)^2\sigma(\omega,T)$ being 
used to define $\tilde\kappa(\omega,T)$.  The chemical potential 
drops out of this formula entirely, hence, one could 
calculate $L(\omega,T)$ in a purely canonical ensemble.  For non-interacting electrons it is 
easy to show that at $T=0$ the Lorenz number is equal to $L_0=(\pi k_B/\sqrt{3}q_e)^2$.  
This result is simply the familiar Wiedemann-Franz law, c.f. Ref.~\onlinecite{am,ziman}.  The 
way this number is obtained in the non-interacting case via our 
formalism is that at $T=0$ there is a delicate balance similar to that in the thermopower, namely, 
\begin{eqnarray}
\lim_{T\rightarrow0}\left(T^2\frac{\tilde\kappa(\omega,T)}{\sigma(\omega,T)}-
\left\{T\frac{\tilde\gamma(\omega,0)}{\sigma(\omega,0)}\right\}^2\right)=0\;.
\end{eqnarray} 
The value of $L_0$ comes from the temperature dependence of both terms.  As
$T\rightarrow0$ both $\tilde\kappa(\omega,T)/T\sigma(\omega,T)$ and 
$(\tilde\gamma(\omega,T)/\sigma(\omega,T))^2$ behave quadratically in 
temperature and the difference in the coefficients multiplying each 
quadratic term is equal precisely to $L_0$. 

\begin{figure}[t]
\begin{center}
\mbox{\bf (a)} 
\includegraphics[width=8.cm]{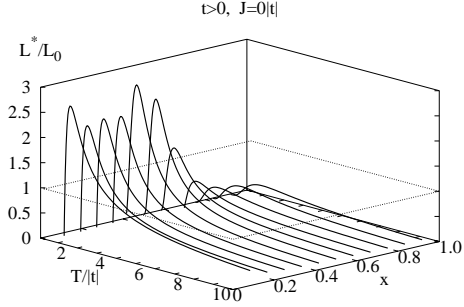}\\
\mbox{\bf (b)} 
\includegraphics[width=8.cm]{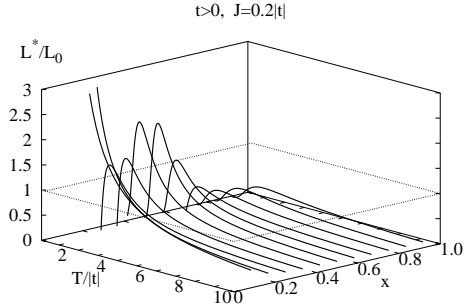}
\end{center}
\caption{$L^*(T)/L_0$ as a function of doping $x$ and temperature $T$ 
for positive hopping $t>0$ (corresponding to NCO after a particle-hole 
transformation) where $L_0=(\pi k_B/\sqrt{3}q_e)^2$.  
Panel {\bf (a)} and {\bf (b)} are for $J=0$ and 
$0.2|t|$, respectively.}
\label{lstar_t1}
\end{figure}

\begin{figure*}[t]
\centering
\begin{tabular}{cc}
\mbox{\bf (a)} 
\includegraphics[width=7.5cm]{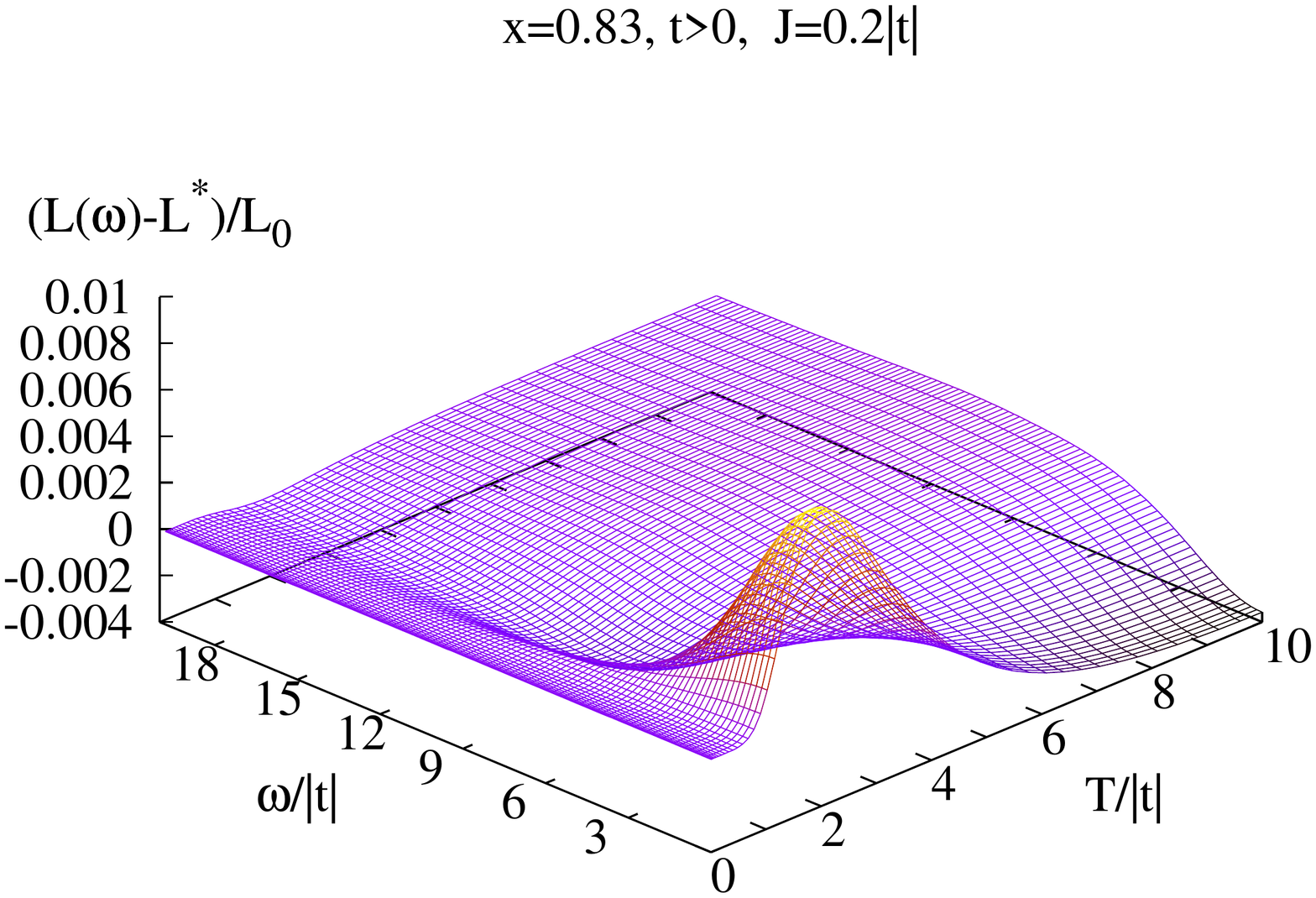}&
\mbox{\bf (b)} 
\includegraphics[width=7.5cm]{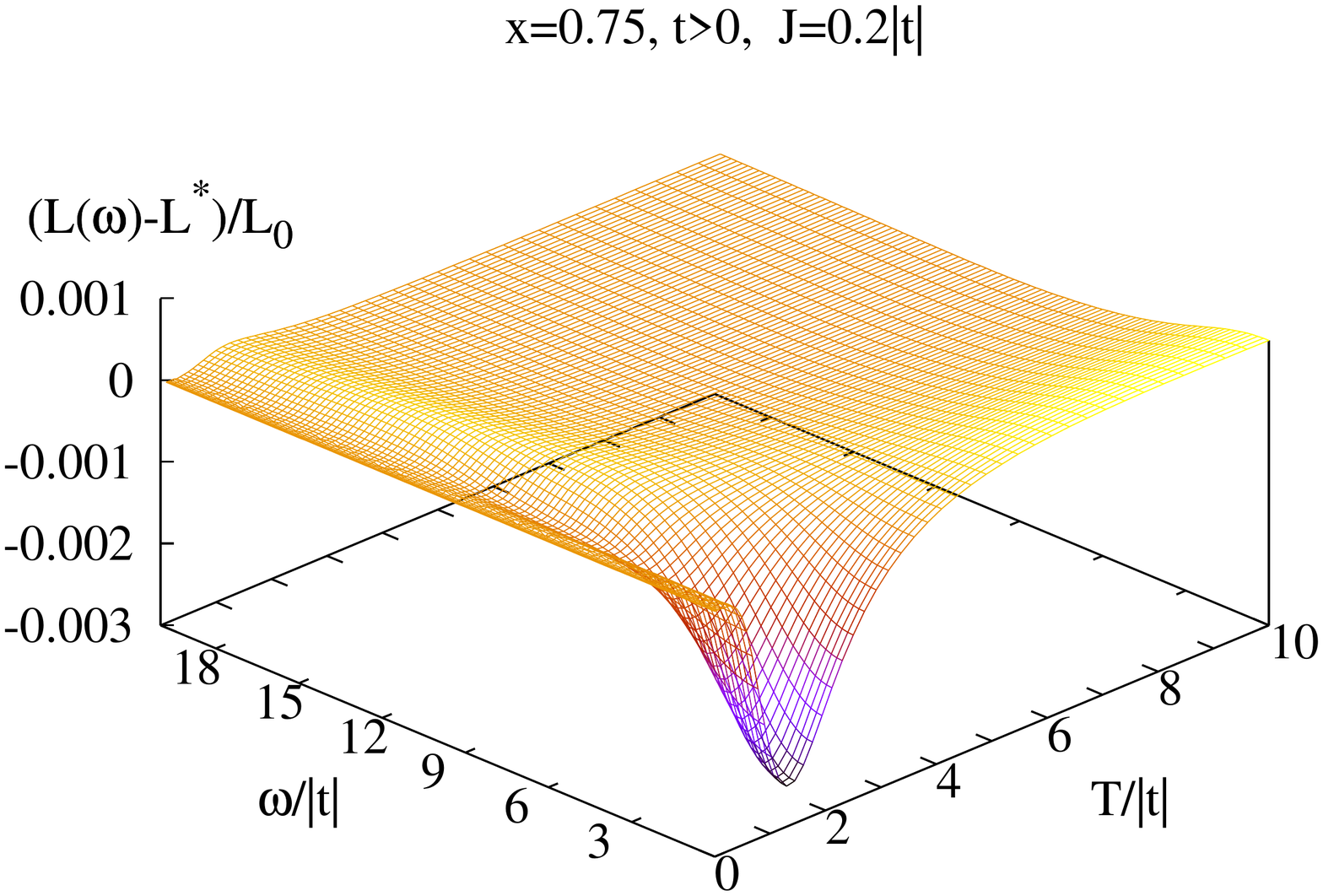}\\
\mbox{\bf (c)} 
\includegraphics[width=7.5cm]{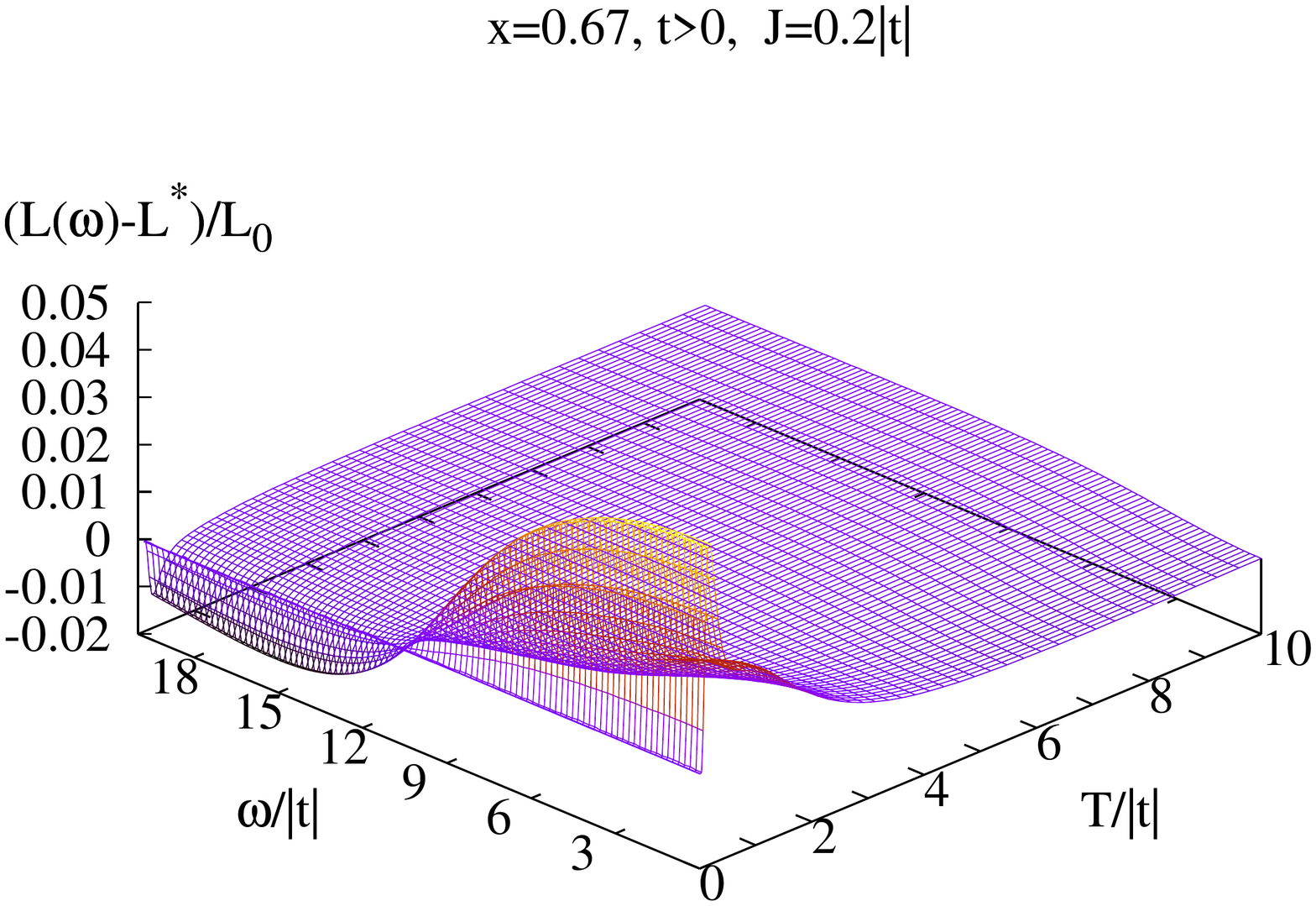}&
\mbox{\bf (d)} 
\includegraphics[width=7.5cm]{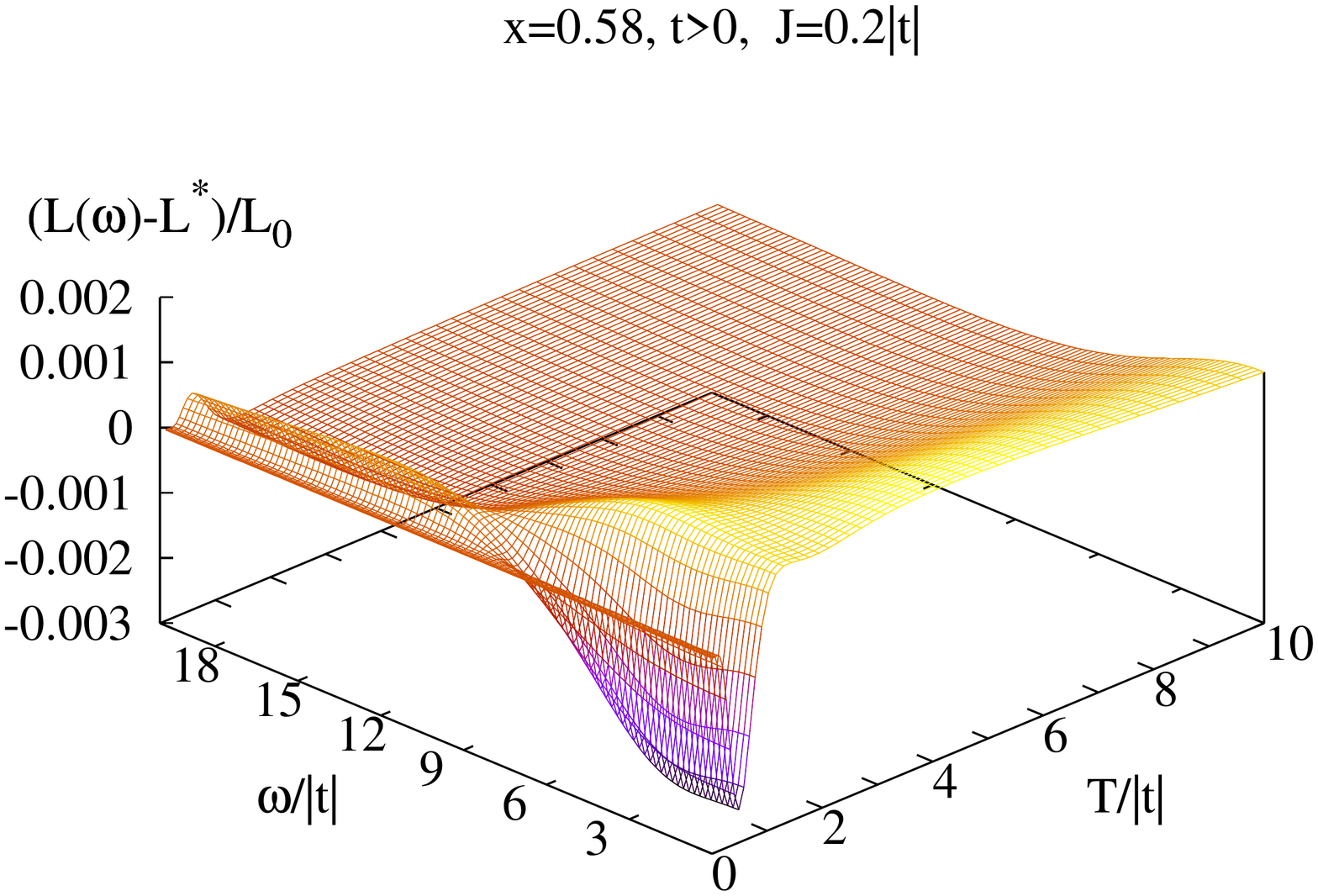}\\
\mbox{\bf (e)} 
\includegraphics[width=7.5cm]{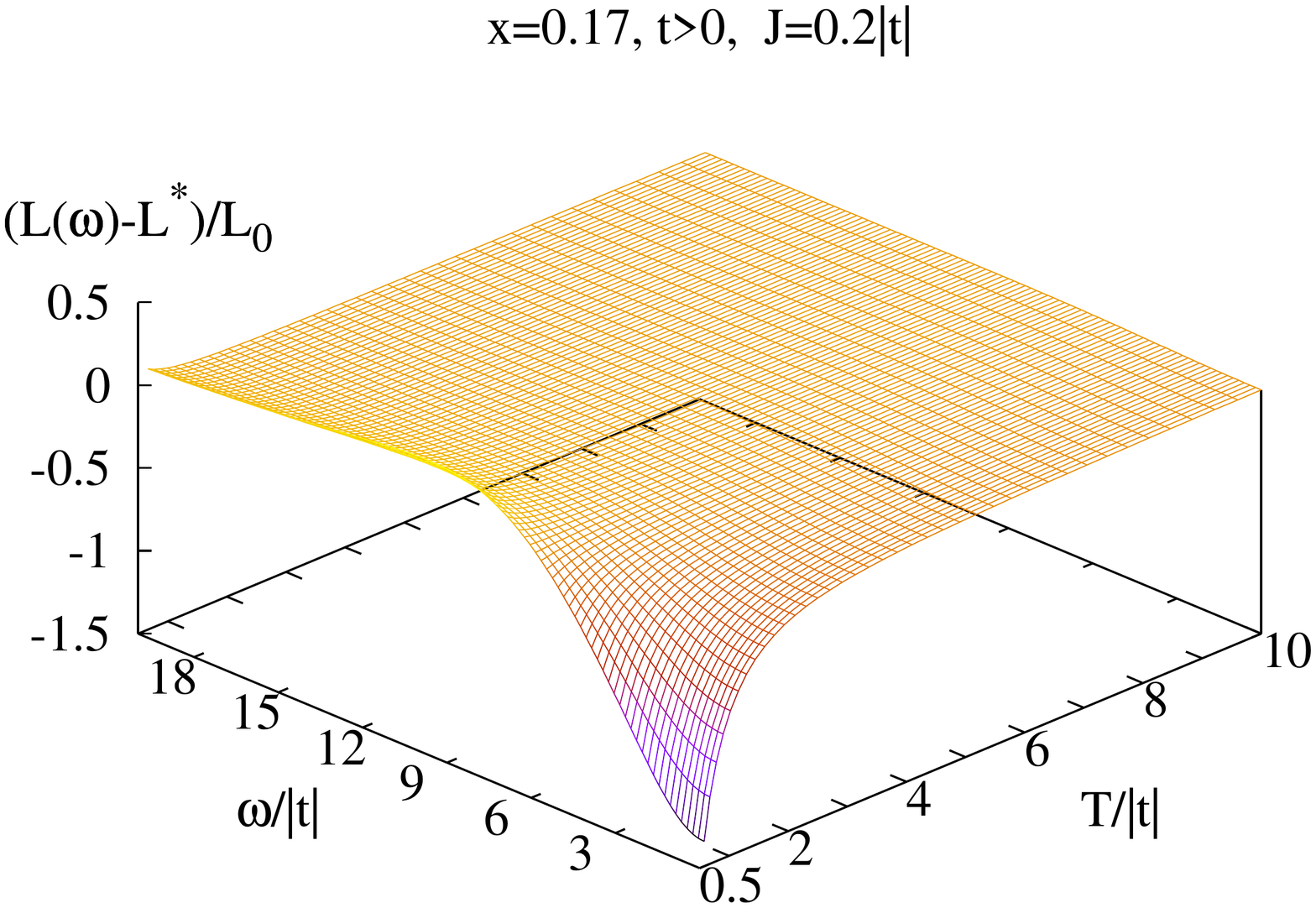}&
\mbox{\bf (f)} 
\includegraphics[width=7.5cm]{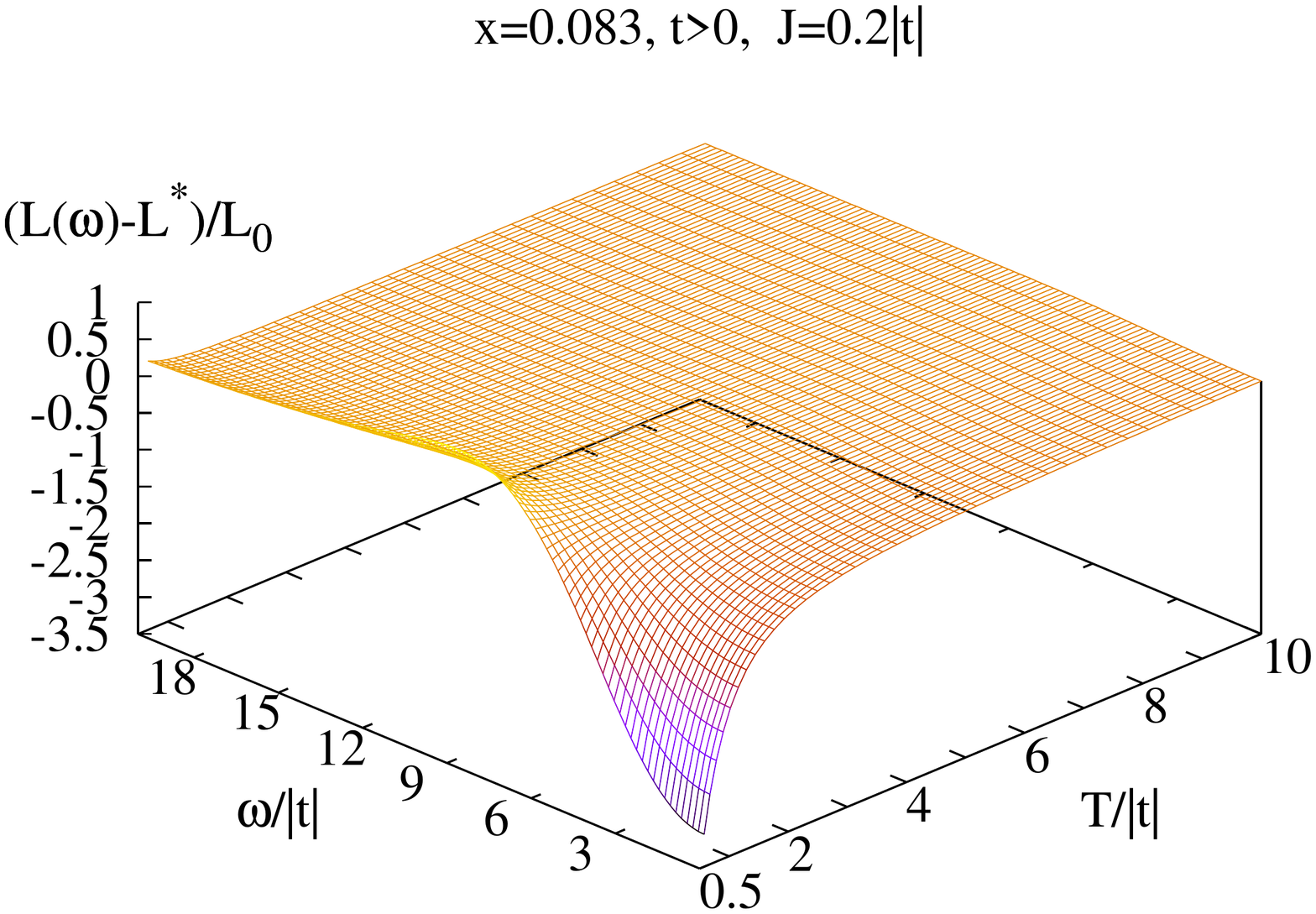}
\end{tabular}
\caption{(color online) $(L(\omega,T)-L^*(T))/L_0$ as a function of frequency $\omega$ 
and temperature $T$ for $J=0.2|t|$, positive hopping $t>0$ (corresponding to NCO after a particle-hole 
transformation), and dopings {\bf (a)} $x=0.83$, {\bf (b)} $x=0.75$, {\bf (c)} $x=0.67$, 
{\bf (d)} $x=0.58$, {\bf (e)} $x=0.17$, and {\bf (f)} $x=0.083$.  The frequency dependence is weak for 
dopings $x\geq0.58$ ({\bf (a)}-{\bf (d)}).  For $x=$0.17, and 0.083 ({\bf (e)} and {\bf (f)}) 
there is strong dependence occurring at small $\omega$ and $T$.  Note 
that in {\bf (e)} and ({\bf (f)} we are only considering 
temperatures $T\geq0.5|t|$ as the Lorenz number badly diverges below that temperature.  
This is most likely a finite size effect and not 
a intrinsic property of the $t$-$J$ model at these dopings.}
\label{l-lstar_t1}
\end{figure*}

In our finite sized system we have neither the delicate balance of the 
$T=0$ behavior nor the quadratic low temperature behavior present.  Using 
the same ``trick'' as we used for the thermopower we force each 
term to separately vanish at $T=0$, i.e., 
\begin{eqnarray}
L(\omega,T)&=&\frac{1}{T^2}\left\{\frac{T\tilde\kappa(\omega,T)}{\sigma(\omega,T)}-
\frac{T\tilde\kappa(\omega,0)}{\sigma(\omega,0)}\right\}\nonumber\\
&&-\frac{1}{T^2}\left\{\left(\frac{T\tilde\gamma(\omega,T)}{\sigma(\omega,T)}\right)^2
-\left(\frac{T\tilde\gamma(\omega,0)}{\sigma(\omega,0)}\right)^2\right\}\;,\nonumber\\
\end{eqnarray} 
to keep $L(\omega,T)$ from 
badly diverging.  At low temperatures the exponential behavior 
caused by the discrete nature of the energy levels of the finite system 
rears its head causing problems as $T\rightarrow0$.

At present we are only able to consider a couple of issues 
regarding the Lorenz number.  Firstly, we can still quite satisfactorily 
determine the accuracy of using the high frequency expansion of $L^*(T)$ in 
place of $L(\omega,T)$.  Secondly, we can determine generally qualitative 
aspects of the Lorenz number for strongly correlated systems as both 
a function of temperature and doping.  Our qualitative 
determination of $L(\omega,T)$ also allows us to look at the 
dimensionless figure of merit which we investigate in 
section~\ref{sec-zt} again at a qualitative level.

The high frequency expansion of $L(\omega,T)$ is 
\begin{eqnarray}
L^*(T)=\frac{\langle\hat{\tilde\Theta}_{xx}\rangle\langle\hat\tau_{xx}\rangle-
\langle\hat{\tilde\Phi}_{xx}\rangle^2}{T^2\langle\hat\tau_{xx}\rangle^2}
\label{lstar}
\end{eqnarray}
where similarly to $\tilde\kappa(\omega,T)$ we have 
defined $\hat{\tilde\Theta}_{xx}$ through $\hat\Theta_{xx}=\hat{\tilde\Theta}_{xx}-(2\mu(T)/q_e)\hat{\tilde\Phi}_{xx} + 
(\mu(T)/q_e)^2\hat\tau_{xx}$.  After forcing each term to vanish at $T=0$ Eq.~\ref{lstar} becomes
\begin{eqnarray}
L^*(T)&=&\frac{1}{T^2}\left\{\frac{\langle\hat{\tilde\Theta}_{xx}(T)\rangle}{\langle\tau_{xx}(T)\rangle}-
\frac{\langle\hat{\tilde\Theta}_{xx}(0)\rangle}{\langle\tau_{xx}(0)\rangle}\right\}\nonumber\\
&&-\frac{1}{T^2}\left\{\frac{\langle\hat{\tilde\Phi}_{xx}(T)\rangle^2}{\langle\tau_{xx}(T)\rangle^2}-
\frac{\langle\hat{\tilde\Phi}_{xx}(0)\rangle^2}{\langle\tau_{xx}(0)\rangle^2}\right\}\;.
\end{eqnarray}

In Figs.~\ref{lstar_t1}a-b and~\ref{l-lstar_t1}a-f we report 
results for $L^*(T)/L_0$, and $(L(\omega,T)-L^*(T))/L_0$ 
for the case of the hopping relevant to NCO.  The 
dotted black line in Fig.~\ref{lstar_t1}a-b is to indicate unity.  It is a reasonable assumption 
that at $T=0$ a thermodynamically large system would have a finite 
value of $L^*(T)$ at $T=0$.  Whether this finite value is equal to $L_0$ is 
an open question and one we are unfortunately not able 
to shed light upon at this time.  

For the non-interacting case the behavior of $L^*(T)$ quickly deviates from $L_0$ 
where it is pinned at $T=0$ as a function of temperature.  For the interacting 
case shown here, the Lorenz number quickly decays to very small values as $T$ increases as
well.  However, our results indicate that the intermediate temperature behavior of $L^*(T)$ generally grows 
with decreasing doping $x$, i.e., interactions evidently increase the 
Lorenz number.  The effect of $J$ is much harder to discern due to presumably 
finite size effects and low temperature divergences.  $J$ seems to have 
little effect except for the highest dopings calculated $x=$0.17, and 0.083.

Fig.~\ref{l-lstar_t1} is similar to Fig.~\ref{s-sstar_t1} and~\ref{s-sstar_t-1}.  For 
dopings $x\geq0.58$ the frequency dependence of $L(\omega,T)$ is very weak and 
only on the order of approximately $\sim(2-3)\%$ or less.  At the low dopings of 
$x=$0.17, and 0.083 there is again a much a stronger frequency dependence that is 
likely due to finite size effects and not intrinsic to the $t$-$J$ model.  

With confidence in the weak frequency dependence of the thermopower 
and the Lorenz number we proceed to calculate the dimensionless FOM for 
the situation applicable to NCO.

\section{Figure of merit}\label{sec-zt}

\begin{figure}[t]
\begin{center}
\mbox{\bf (a)} 
\includegraphics[width=8.cm]{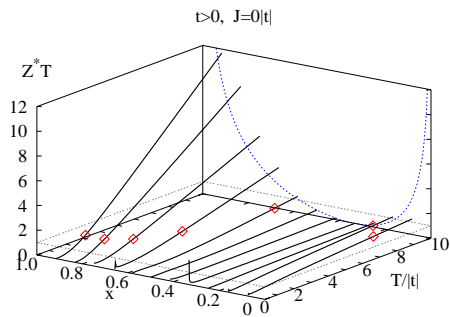}\\
\mbox{\bf (b)} 
\includegraphics[width=8.cm]{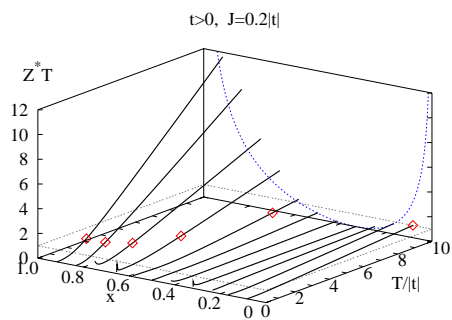}
\end{center}
\caption{(color online) $Z^*(T)T$ as a function of doping $x$ and 
temperature $T$ for {\bf (a)} $J=0$ and {\bf (b)} $0.2|t|$.  The 
red squares indicate the point at which $Z^*(T)T=1$ while the black dotted 
line indicates unity as well.}
\label{zstart_t1}
\end{figure}

A value of the dimensionless figure of merit greater than one 
is indicative of a good thermoelectric material and is therefore desired.  
Of course, throughout this work we have neglected the 
lattice contribution to the Lorenz number 
which will add to our calculation of $L(\omega,T)$ and consequently 
serve to decrease the FOM that we calculate below.

Recall that the FOM is given in Eq.~\ref{fom} while the high frequency 
expansion is given by Eq.~\ref{fom-hfe}.  The numerator is just the 
square of the thermopower and vanishes at zero temperature and eventually 
obtains the square of the MH limit, i.e., $[(86\mu V/K)\ln(2x/(1-x))]^2$.  However, 
the denominator is the Lorenz number which starts out finite at $T=0$ and quickly 
decays to zero as $T\rightarrow\infty$.  Therefore we expect the FOM to begin at 
zero and grow without bound as $T$ increases.  

Fig.~\ref{zstart_t1}a-b shows the FOM for $J=0$ and $J=0.2|t|$ for the 
positive case of the hopping $t$ as a function of both the doping and 
temperature.  True to expectations the FOM quickly grows from zero at 
$T=0$ to well above unity for dopings $x\geq0.58$.  At intermediate 
dopings ($0.5\geq x\geq0.17$) the FOM remains very small up to the 
largest temperatures calculated and never reaches unity.  This is understandable 
when one considers that the thermopower in this regime has a small absolute 
value for all temperatures 
and the Lorenz number only serves to diminish this value.  At the 
smallest doping calculated ($x=0.083$) the FOM just reaches 
unity as $T\sim10|t|$.  

Note that in Fig.~\ref{zstart_t1} we have indicated 
the point at which $Z^*(T)T$ equals one by a red box.  Projected onto 
the $T=10|t|$ plane is the ``normalized'' square of the 
MH term, i.e., $(\ln(2x/(1-x)))^2$ indicating an interesting doping 
behavior of the the relative magnitudes of the figure of merit.  This behavior is 
most certainly due to the MH term of the thermopower.  Nonzero $J$ 
has a nearly negligible effect accept at the lowest dopings where 
it serves to slightly reduce the highest temperature value of the FOM.

The full frequency dependence of $Z(\omega,T)T$ is understandably weak (not shown) 
since both the thermopower and Lorenz number have been shown to be weakly 
frequency dependent.

\section{Conclusion}\label{sec-conc}

In this work we have established the general validity of the high frequency 
expansion of the thermopower, Lorenz number, and figure of merit for 
strongly correlated electron models.  This high frequency expansion is much 
simpler to consider than the full Kubo formalism and yet complicated enough to 
capture the full interaction effects.  This established validity should provide 
a benchmark encouraging the theoretical community to obtain useful approximate 
methods to calculate these high frequency formulas.

We also provide theoretical evidence supporting the authors' previous calculations 
in regard to the Curie-Weiss metallic phase of NCO\cite{cw-prl,cw-prb} 
being qualitatively and quantitatively described by a two-dimensional $t$-$J$ model 
on a triangular lattice.  Further, we have provided new predictions for sodium 
cobalt oxide concerning the Lorenz number and figure of merit (Secs.~\ref{sec-l} and~\ref{sec-zt}).  
In Ref.~\onlinecite{cw-prl,cw-prb} the present authors used an experimental hopping 
parameter of $|t|\sim 100 K$ taken from both photoemission 
experiments\cite{hasan,yang_1} and chemical potential 
measurements\cite{fujimori} indicating NCO to have a
very narrow band system.  Therefore, one should scale the temperature in this 
work by this value of the hopping when making experimental comparisons.

The predicted thermopower enhancement discussed in Sec.~\ref{s_neg_t} should stimulate an 
experimental search of lattice based strongly correlated materials to provide 
useful thermolectric materials.  Perhaps it is possible to custom design 
a high thermopower material armed with the knowledge of the behavior of the 
transport term alone for strongly correlated systems.

\begin{acknowledgments}
We gratefully acknowledge support from Grants NSF-DMR0408247 and 
DOE BES DE-FG02-06ER46319.  We also thank Prof. J. B. Anderson (Chemistry) and Eric 
Prescott (Computer Science and Engineering) 
at the Pennsylvania State University for allowing time on the 
computer cluster MUFASA which is supported by CEMBA, an NSF funded IGERT program.  
We also acknowledge enlightening conversations with Subroto Mukerjee.
\end{acknowledgments}

\begin{widetext}
\appendix*
\section{Operator Expressions}

This appendix presents formulas taken from Ref.~\onlinecite{shastry_1,shastry_2,shastry_3} for completeness and
clarity and additional ones used in the present work.

In discussing the explicit forms of the operators we introduce the so-called 
Hubbard operators to act to simplify the formulas somewhat.  The Hubbard operators 
are defined as follows:
\begin{eqnarray}
X_{\sigma 0}(\vec r) &\equiv& \tilde c^{\dagger}_{\vec r\sigma}\\
X_{0\sigma}(\vec r) &\equiv& \tilde c_{\vec r\sigma}\\
X_{\sigma\sigma'}(\vec r) &\equiv& \tilde c^{\dagger}_{\vec r\sigma}\tilde
c_{\vec r\sigma'} = X_{\sigma 0}(\vec r)X_{0\sigma'}(\vec r)\\
X_{00}(\vec r) &\equiv & \tilde c_{\vec r\sigma}\tilde c^{\dagger}_{\vec
r\sigma}= X_{0\sigma}(\vec r)X_{\sigma 0}(\vec r)\;.
\end{eqnarray}
They have modified anti-commutation relations
\begin{eqnarray}
\{\tilde c^{\dagger}_{\vec r\sigma},\tilde c_{\vec r'\sigma'}\}
&=&\{X_{\sigma0}(\vec r),X_{0\sigma}(\vec r')\}\nonumber\\
&=&\delta_{\vec r\vec r'}(X_{\sigma\sigma'}(\vec
r')+\delta_{\sigma\sigma'}X_{00}(\vec r'))\nonumber\\
&\equiv&\delta_{\vec r\vec r'}Y_{\sigma'\sigma}(\vec r')
\end{eqnarray}
where the last line defines $Y_{\sigma'\sigma}(\vec r)$.  
Using this notation the charge and energy currents defined 
in Eq.~\ref{j} and Eq.~\ref{je} are given by
\begin{eqnarray}
\hat{J}_x = -\lim_{k_x\rightarrow0}\frac{d}{dk_x}[\hat K(k_x),q_e\hat{n}(-k_x)] = iq_et\sum_{r\eta\sigma}\eta_x X_{\sigma0}(\vec r+\vec\eta)X_{0\sigma}(\vec r)\;,
\end{eqnarray}
and
\begin{eqnarray}
\hat J_x^E &=& -\lim_{k_x\rightarrow0}\frac{d}{dk_x}[\hat T(k_x),\frac{1}{2}\hat
T(-k_x) + \hat U(-k_x)]\nonumber\\
 &=& -\frac{it^2}{2}\sum_{r\eta\eta'\sigma\sigma'}(\eta_x+\eta'_x)
Y_{\sigma\sigma'}(\vec r+\vec\eta)
X_{\sigma0}(\vec r+\vec\eta+\vec\eta')X_{0\sigma'}(\vec r)
+\frac{iJt}{4}\sum_{r\eta\eta'\sigma}\{\eta_x\vec\mu(\vec r+\vec\eta,\vec r)
\cdot[\vec S_{\vec r+\vec\eta'} + 
\vec S_{\vec r+\vec\eta+\vec\eta'}]\nonumber\\
&&+[(\eta_x+2\eta'_x)\vec S_{\vec r+\vec\eta+\vec\eta'} + 
(\eta_x-2\eta'_x)\vec S_{\vec r+\vec\eta'}]\cdot \vec\mu(\vec r+\vec\eta,\vec r)\}\;,
\end{eqnarray}
respectively.

The stress tensor is simply
\begin{eqnarray}
\hat\tau_{xx} = -\lim_{k_x\rightarrow0}\frac{d}{dk_x}[\hat{J}_x(k_x),\hat
q_en(-k_x)] = q_e^2t\sum_{r\eta\sigma}\eta_x^2 X_{\sigma0}(\vec r+\vec\eta)X_{0\sigma}(\vec r)\;,
\end{eqnarray}
while the modified  thermoelectric, and thermal 
operators ($\hat{\tilde\Phi}_{xx}$ and $\hat{\tilde\Theta}_{xx}$) have the form
\begin{eqnarray}
\hat{\tilde\Phi}_{xx} &=& -\lim_{k_x\rightarrow0}\frac{d}{dk_x}[\hat{J}_x(k_x),\hat{H}(-k_x)]\nonumber\\
 &=& -q_e\frac{t^2}{2}\sum_{r\eta\eta'\sigma\sigma'}
(\eta_x+\eta'_x)^2Y_{\sigma\sigma'}(\vec r+\vec\eta)
X_{\sigma0}(\vec r+\vec\eta+\vec\eta')X_{0\sigma'}(\vec r)
+\frac{q_etJ}{4}\sum_{r\eta\eta'\sigma}\{\eta_x[\eta_x \vec\mu(\vec r+\vec\eta,\vec r)
\cdot [\vec S_{\vec r+\vec\eta'} + 
\vec S_{\vec r+\vec\eta+\vec\eta'}]\nonumber\\
&&+[(\eta_x+2\eta'_x)\vec S_{\vec r+\vec\eta+\vec\eta'} + 
(\eta_x-2\eta'_x)\vec S_{\vec r+\vec\eta'}]\cdot \vec\mu(\vec r+\vec\eta,\vec r)\}\;,
\end{eqnarray}
and
\begin{eqnarray}\label{theta-tilde}
\hat{\tilde\Theta}_{xx} &=& -\lim_{k_x\rightarrow0}\frac{d}{dk_x}[\hat{J}_x^E(k_x),\hat{H}(-k_x)]\nonumber\\
 &=& \frac{t^3}{4}\sum_{r\eta\eta'\eta''}\{\sum_{\sigma}(\eta_x+\eta'_x)(-\eta_x+\eta'_x-\eta''_x)
[\{X_{\bar\sigma0}(\vec r+\vec\eta)X_{0\sigma}(\vec r+\vec\eta+\vec\eta'')-\nonumber\\&&
X_{\bar\sigma0}(\vec r+\vec\eta+\vec\eta'')X_{0\sigma}(\vec r+\vec\eta)\}
X_{\sigma0}(\vec r+\vec\eta+\vec\eta')X_{0\bar\sigma}(\vec r)\nonumber\\&&
-\{X_{\bar\sigma0}(\vec r+\vec\eta)X_{0\bar\sigma}(\vec r+\vec\eta+\vec\eta'')-h.c.\}X_{\sigma0}(\vec r+\vec\eta+\vec\eta'')X_{0\sigma}(\vec r)]\nonumber\\
&&+\sum_{\sigma\sigma'\sigma''}(\eta_x+\eta'_x+\eta''_x)(\eta_x+2\eta'_x+\eta''_x)
Y_{\sigma\sigma'}(\vec r+\vec\eta)Y_{\sigma''\sigma'}(\vec r+\vec\eta+\vec\eta')X_{\sigma'''0}(\vec r+\vec\eta+\vec\eta'+\vec\eta'')X_{0\sigma'}(\vec r)\}\nonumber\\
&&+\frac{t^2J}{16}\sum_{r\eta\eta'\eta''}(\eta_x+\eta'_x)\{\sum_{\sigma\sigma'}
[(\eta_x-\eta'_x)X_{\sigma0}(\vec r+\vec\eta+\vec\eta')X_{0\sigma'}(\vec r)[X_{\sigma'\bar\sigma}(\vec r+\vec\eta)X_{\bar\sigma\sigma}(\vec r+\vec\eta+\vec\eta'')\nonumber\\&&
-X_{\bar\sigma'\sigma}(\vec r+\vec\eta)X_{\sigma'\bar\sigma'}(\vec r+\vec\eta+\vec\eta'')]
-(\eta_x+\eta'_x)Y_{\sigma\sigma'}(\vec r+\vec\eta)X_{\bar\sigma0}(\vec r+\vec\eta+\vec\eta')X_{0\sigma'}(\vec r)
X_{\sigma\bar\sigma}(\vec r+\vec\eta+\vec\eta'+\vec\eta'')\nonumber\\
&&-(\eta_x+\eta'_x)Y_{\sigma\sigma'}(\vec r+\vec\eta)X_{\sigma0}(\vec r+\vec\eta+\vec\eta')X_{0\bar\sigma'}(\vec r)
X_{\bar\sigma'\sigma}'(\vec r+\vec\eta'')\nonumber\\
&&+(\eta_x-\eta'_x+2\eta''_x)[X_{\bar\sigma\sigma}(\vec r+\vec\eta+\vec\eta'')X_{\sigma'\bar\sigma}(\vec r+\vec\eta)
-X_{\sigma'\bar\sigma}(\vec r+\vec\eta+\vec\eta'')X_{\bar\sigma'\sigma}(\vec r+\vec\eta)]
X_{\sigma0}(\vec r+\vec\eta+\vec\eta')X_{0\sigma'}(\vec r)\nonumber\\
&&-(\eta_x+\eta'_x+2\eta''_x)X_{\sigma\bar\sigma}(\vec r+\vec\eta+\vec\eta'+\vec\eta'')Y_{\sigma\sigma'}(\vec r+\vec\eta)
X_{\bar\sigma0}(\vec r+\vec\eta+\vec\eta')X_{0\sigma'}(\vec r)\nonumber\\
&&-(\eta_x+\eta'_x-2\eta''_x)X_{\bar\sigma'\sigma'}(\vec r+\vec\eta'')Y_{\sigma\sigma'}(\vec r+\vec\eta)
X_{\sigma0}(\vec r+\vec\eta+\vec\eta')X_{0\bar\sigma'}(\vec r)\nonumber\\
&&-(\eta_x+\eta'_x-\eta''_x)Y_{\sigma\sigma'}(\vec r+\vec\eta)X_{\sigma0}(\vec r+\vec\eta+\vec\eta')X_{0\sigma'}(\vec r)
[n_{\vec r+\vec\eta''\sigma'}-n_{\vec r+\vec\eta''\bar\sigma'}]\nonumber\\
&&-(\eta_x+\eta'_x+\eta''_x)Y_{\sigma\sigma'}(\vec r+\vec\eta)[n_{\vec r+\vec\eta+\vec\eta'+\vec\eta''\sigma}-n_{\vec r+\vec\eta+\vec\eta'+\vec\eta''\bar\sigma}]
X_{\sigma0}(\vec r+\vec\eta+\vec\eta')X_{0\sigma'}(\vec r)]\nonumber\\
&&+(2\eta_x-2\eta'_x+2\eta''_x)\sum_{\sigma}[n_{\vec r+\vec\eta+\vec\eta''\sigma}-n_{\vec r+\vec\eta+\vec\eta''\bar\sigma}]
X_{\bar\sigma\sigma}(\vec r+\vec\eta)X_{\sigma0}(\vec r+\vec\eta+\vec\eta')X_{0\bar\sigma}(\vec)\}\nonumber\\
&&+\frac{tJ^2}{8}\sum_{r\eta\eta'\eta''}\{\sum_{\alpha\beta\gamma}i\epsilon_{\alpha\beta\gamma}[\eta_x\mu^{\alpha}(\vec r+\vec\eta,\vec r)
\{S^\beta_{\vec r+\vec\eta''}\left[\frac{\eta_x}{4}S^\gamma_{\vec r+\vec\eta''}-\frac{\eta_x}{4}S^\gamma_{\vec r+\vec\eta+\vec\eta''}-\frac{(\eta_x-2\eta'_x)}{2}S^\gamma_{\vec r+\vec\eta'+\vec\eta''}\right]\nonumber\\
&&+S^\beta_{\vec r+\vec\eta+\vec\eta'}\left[\frac{\eta_x}{4}S^\gamma_{\vec r+\vec\eta''}-\frac{\eta_x}{4}S^\gamma_{\vec r+\vec\eta+\vec\eta''}+\frac{(\eta_x+2\eta'_x)}{2}S^\gamma_{\vec r+\vec\eta+\vec\eta'+\vec\eta''}\right]\}\nonumber\\
&&+\eta_x\left[\frac{(\eta_x-2\eta''_x)}{4}S^\alpha_{\vec r+\vec\eta''}-\frac{(\eta_x+2\eta''_x)}{4}S^\alpha_{\vec r+\vec\eta+\vec\eta''}-\frac{(\eta_x-2\eta'_x-2\eta''_x)}{2}S^\alpha_{\vec r+\vec\eta'+\vec\eta''}\right]\mu^\beta(\vec r+\vec\eta,\vec r)S^\gamma_{\vec r+\vec\eta'}\nonumber\\
&&+\eta_x\left[\frac{(\eta_x-2\eta''_x)}{4}S^\alpha_{\vec r+\vec\eta''}-\frac{(\eta_x+2\eta''_x)}{4}S^\alpha_{\vec r+\vec\eta+\vec\eta''}+\frac{(\eta_x+2\eta'_x+2\eta''_x)}{2}S^\alpha_{\vec r+\vec\eta+\vec\eta'+\vec\eta''}\right]\mu^\beta(\vec r+\vec\eta,\vec r)S^\gamma_{\vec r+\vec\eta+\vec\eta'}\nonumber\\
&&+(\eta_x+2\eta'_x)[S^\alpha_{\vec r+\vec\eta+\vec\eta'}\mu^\beta(\vec r+\vec\eta,\vec r)\left\{\frac{\eta_x}{4}S^\gamma_{\vec r+\vec\eta+\vec\eta''}-\frac{\eta_x}{4}S^\gamma_{\vec r+\vec\eta''}-\frac{(\eta_x+2\eta'_x)}{2}S^\gamma_{\vec r+\vec\eta+\vec\eta'+\vec\eta''}\right\}\nonumber\\
&&+\left\{\frac{(\eta_x+2\eta''_x)}{4}S^\alpha_{\vec r+\vec\eta+\vec\eta''}-\frac{(\eta_x-2\eta''_x)}{4}S^\alpha_{\vec r+\vec\eta''}-\frac{(\eta_x+2\eta'_x+2\eta''_x)}{2}S^\alpha_{\vec r+\vec\eta+\vec\eta'+\vec\eta''}\right\}S^\beta_{\vec r+\vec\eta+\vec\eta'}\mu^\gamma(\vec r+\vec\eta,\vec r)]\nonumber\\
&&+(\eta_x-2\eta''_x)[S^\alpha_{\vec r+\vec\eta'}\mu^\beta(\vec r+\vec\eta,\vec r)\left\{\frac{\eta_x}{4}S^\gamma_{\vec r+\vec\eta+\vec\eta''}-\frac{\eta_x}{4}S^\gamma_{\vec r+\vec\eta''}+\frac{(\eta_x-2\eta'_x)}{2}S^\gamma_{\vec r+\vec\eta'+\vec\eta''}\right\}\nonumber\\
&&+\left\{\frac{(\eta_x+2\eta''_x)}{4}S^\alpha_{\vec r+\vec\eta+\vec\eta''}-\frac{(\eta_x-2\eta''_x)}{4}S^\alpha_{\vec r+\vec\eta''}+\frac{(\eta_x-2\eta'_x-2\eta''_x)}{2}S^\alpha_{\vec r+\vec\eta'+\vec\eta''}\right\}S^\beta_{\vec r+\vec\eta'}\mu^\gamma(\vec r+\vec\eta,\vec r)]
]\nonumber\\
+&&\sum_{\alpha}[\frac{\eta^2_x}{8} X_{\sigma0}(\vec r+\vec\eta)X_{0\sigma}(\vec r)\left\{S^\alpha_{\vec r+\vec\eta'}+S^\alpha_{\vec r+\vec\eta+\vec\eta'}\right\}\left\{S^\alpha_{\vec r+\vec\eta''}+S^\alpha_{\vec r+\vec\eta+\vec\eta''}\right\}\nonumber\\
&&+\left\{\frac{\eta_x(\eta_x-2\eta''_x)}{8}S^\alpha_{\vec r+\vec\eta''}+\frac{\eta_x(\eta_x+2\eta_x'')}{8}S^\alpha_{\vec r+\vec\eta+\vec\eta''}\right\}X_{\sigma0}(\vec r+\vec\eta)X_{0\sigma}(\vec r)\left\{S^\alpha_{\vec r+\vec\eta'}+S^\alpha_{\vec r+\vec\eta+\vec\eta'}\right\}\nonumber\\
&&+\frac{(\eta_x+2\eta'_x)}{8}\{S^\alpha_{\vec r+\vec\eta+\vec\eta'}X_{\sigma0}(\vec r+\vec\eta)X_{0\sigma}(\vec r)[\eta_x S^\alpha_{\vec r+\vec\eta''}+\eta_x S^\alpha_{\vec r+\vec\eta+\vec\eta''}]\nonumber\\
&&+[(\eta_x-2\eta''_x)S^\alpha_{\vec r+\vec\eta''}+(\eta_x+2\eta''_x)S^\alpha_{\vec r+\vec\eta+\vec\eta''}]S^\alpha_{\vec r+\vec\eta+\vec\eta'}X_{\sigma0}(\vec r+\vec\eta)X_{0\sigma}(\vec r)\}\nonumber\\
&&+\frac{(\eta_x-2\eta'_x)}{8}\{S^\alpha_{\vec r+\vec\eta'}X_{\sigma0}(\vec r+\vec\eta)X_{0\sigma}(\vec r)[\eta_xS^\alpha_{\vec r+\vec\eta''}+\eta_xS^\alpha_{\vec r+\vec\eta+\vec\eta''}]\nonumber\\
&&+[(\eta_x-2\eta''_x)S^\alpha_{\vec r+\vec\eta''}+(\eta_x+2\eta''_x)S^\alpha_{\vec r+\vec\eta+\vec\eta''}]S^\alpha_{\vec r+\vec\eta'}X_{\sigma0}(\vec r+\vec\eta)X_{0\sigma}(\vec r)\}
]\}\;,
\end{eqnarray}
respectively.  In the above expression for $\hat{\tilde\Phi}_{xx}$ and $\hat{\tilde\Theta}_{xx}$ we have 
used the bond spin operators defined as
\begin{eqnarray}
&&\mu^z(\vec r,\vec r')=\frac{1}{2}[X_{\uparrow 0}(\vec r)X_{0\uparrow}(\vec r') - X_{\downarrow 0}(\vec r)X_{0\downarrow}(\vec r')]\\
&&\mu^y(\vec r,\vec r')=\frac{1}{2i}[X_{\uparrow 0}(\vec r)X_{0\downarrow}(\vec r') - X_{\downarrow 0}(\vec r)X_{0\uparrow}(\vec r')]\\
&&\mu^x(\vec r,\vec r')=\frac{1}{2}[X_{\uparrow 0}(\vec r)X_{0\downarrow}(\vec r') + X_{\downarrow 0}(\vec r)X_{0\uparrow}(\vec r')]\;.
\end{eqnarray}
Further note that for Eq.~\ref{theta-tilde} we have used the definition 
\begin{eqnarray}
\hat{J}^E_x(k_x)=-\lim_{q_x\rightarrow 0}\frac{d}{dq_x}\left\{\frac{1}{2}[\hat{T}(q_x+k_x),\hat{T}(-q_x)]
+[\hat{T}(q_x+k_x),\hat{V}(-q_x)]\right\}
\end{eqnarray}
allowing the establishment of the identity (using the Jacobi identity)
\begin{eqnarray}
\lim_{k_x,q_x\rightarrow0}\frac{d^2}{dk_xdq_x}\left\{[[\hat{T}(k_x+q_x),\hat{V}(-q_x)],\hat{T}(-k_x)]
-\frac{1}{2}[[\hat{T}(k_x+q_x),\hat{T}(-q_x)],\hat{V}(-k_x)]\right\}=0
\end{eqnarray}
simplifying the calculation of Eq.~\ref{theta-tilde} tremendously.

\end{widetext}


\begin{thebibliography}{32}
\bibitem[{T_U()}]{T_U_note}
\bibinfo{note}{That is, the Hamiltonian is written as $\hat{H}=\hat{T}+\hat{U}$
  where $\hat{T} = -\sum_{\vec r\vec\eta\sigma}t(\vec\eta)\tilde
  c^{\dagger}_{\vec r+\vec\eta\sigma} \tilde c_{\vec r\sigma}$ and $\hat{U} =
  \frac{1}{2}\sum_{\vec r\vec \eta}J(\vec \eta)\vec{S}_{\vec r}\cdot
  \vec{S}_{\vec r + \vec \eta}$.}
\bibitem[{mu_()}]{mu_note}
\bibinfo{note}{For a $t$-$J$ model with $\mathcal{L}$ sites the possible number
  of electrons is $N=1\ldots\mathcal{L}-1$. So while
  $\mu(T)=(F_{N+1}-F_{N-1})/2$ is used for $N=2\ldots\mathcal{L}-2$ we define
  it as $\mu(T)=F_{N+1}-F_N$ for $N=2$ and $\mu(T)=F_N-F_{N-1}$ for
  $N=\mathcal{L}-1$.}
\bibitem[{mh_()}]{mh_note}
\bibinfo{note}{It should be said that the Mott-Heikes limit for an uncorrelated
  band of spinfull particles is equivalent to that of the Hubbard model for
  arbitrary $U$. When the $T\rightarrow\infty$ limit is taken the value of $U$
  is essentially scaled to zero compared to the temperature yielding the same
  result as that for an uncorrelated band.}
\end{thebibliography}
\end{document}